%% file: main.tex
\documentclass[acmsmall,screen,nonacm]{acmart}

\setcopyright{none}

\usepackage{booktabs}
\usepackage{subcaption}

\usepackage{program}
\usepackage{centernot}
\usepackage{graphicx}
\usepackage{amsmath}

\usepackage{xparse}
\usepackage{stmaryrd}
\usepackage{here}
\usepackage{mleftright}
\usepackage{color}
\usepackage{multicol}
\usepackage{multirow}
\usepackage{proof}
\usepackage[scaled=0.8]{beramono}
\usepackage[T1]{fontenc}
\usepackage[all]{xy}
\usepackage{pifont}
\usepackage[normalem]{ulem}
\usepackage{algpseudocode}
\usepackage{alltt}
\usepackage{macros}

\begin{document}

\title{Program Verification via Predicate Constraint Satisfiability Modulo Theories}

\author{Hiroshi Unno}
\affiliation{
  \institution{University of Tsukuba}
  \country{Japan}
}
\affiliation{
  \institution{RIKEN AIP}
  \country{Japan}
}
\email{uhiro@cs.tsukuba.ac.jp}

\author{Yuki Satake}
\affiliation{
  \institution{University of Tsukuba}
  \country{Japan}
}
\email{satake@logic.cs.tsukuba.ac.jp}

\author{Tachio Terauchi}
\affiliation{
  \institution{Waseda University}
  \country{Japan}
}
\email{terauchi@waseda.jp}

\author{Eric Koskinen}
\affiliation{
  \institution{Stevens Institute of Technology}
  \country{USA}
}
\email{eric.koskinen@stevens.edu}

\input{abst}
\maketitle

\section{Introduction}
\label{sec:intro}
\input{intro}

\section{Overview}
\label{sec:overview}
\input{overview}

\section{Extended Constraint Logic Programs~\MuCLP{}}
\label{sec:muclp}
\input{muclp}

\section{Predicate Constraint Satisfaction Problems \PCSPWFFN{}}
\label{sec:pcsp}
\input{pcsp}

\section{Reduction Algorithm from \MuCLP{} to \PCSPWFFN{}}
\label{sec:reduct}
\input{reduct}

\section{Constraint Solving Method for \PCSPWFFN{}}
\label{sec:csolve}
\input{csolve}

\section{Evaluation}
\label{sec:eval}
\input{eval}

\section{Related Work}
\label{sec:related}
\input{related}

\section{Conclusion}
\label{sec:conc}
\input{conc}

\input{main.bbl}

\appendix

\section{Expressing Verification Problems in~\MuCLP{}}
\label{sec:apps}
\input{apps}

\end{document}

%% file: abst.tex
\begin{abstract}
This paper presents a verification framework based on a new class of predicate Constraint Satisfaction Problems called \PCSPWFFN{} where constraints are represented as clauses modulo first-order theories over \emph{function variables} and \emph{predicate variables that may represent well-founded predicates}.  The verification framework generalizes an existing one based on Constrained Horn Clauses (\CHCS{}) to arbitrary clauses, function variables, and well-foundedness constraints.  While it is known that the satisfiability of \CHCS{} and the validity of queries for Constrained Logic Programs (\CLP{}) are inter-reducible, we show that, thanks to the added expressiveness, \PCSPWFFN{} is expressive enough to express \MuCLP{} queries.  \MuCLP{} itself is a new extension of \CLP{} that we propose in this paper.  It extends \CLP{} with arbitrarily nested inductive and co-inductive predicates and is equi-expressive as first-order fixpoint logic.  We show that \MuCLP{} can naturally encode a wide variety of verification problems including but not limited to termination/non-termination verification and even full modal mu-calculus model checking of programs written in various languages.  
To establish our verification framework, we present (1) a sound and complete reduction algorithm from \MuCLP{} to \PCSPWFFN{} and (2) a constraint solving method for \PCSPWFFN{} based on \emph{stratified} CounterExample-Guided Inductive Synthesis (CEGIS) of (co-)inductive invariants, ranking functions, and Skolem functions witnessing existential quantifiers.  Stratified CEGIS combines CEGIS with stratified families of templates to achieve relative completeness and faster and stable convergence of CEGIS by avoiding the overfitting problem.  We have implemented the proposed framework and obtained promising results on diverse verification problems that are beyond the scope of the previous verification frameworks based on \CHCS{}.
\end{abstract}

%% file: intro.tex
In the formal verification community, a class of predicate constraints called Constrained Horn Clauses (\CHCS{})~\cite{Bjorner2015a} has been widely adopted as a ``common intermediate language'' for uniformly expressing verification problems for various programming paradigms, such as functional and object-oriented languages.  Example uses of the \CHCS{} framework include safety property verification~\cite{Grebenshchikov2012,Gurfinkel2015,Kahsai2016} and refinement type inference~\cite{Unno2009,Terauchi2010,Kobayashi2011b,Jhala2011,Zhu2015}.  The wide applicability of \CHCS{} is due in no small part to its expressiveness: it is known that the satisfiability of \CHCS{} and the validity of queries for Constrained Logic Programs (\CLP{})~\cite{Jaffar1994} are inter-reducible.
The separation of constraint generation and solving has facilitated the rapid development of constraint generation tools such as \rcaml~\cite{Unno2009}, \seahorn~\cite{Gurfinkel2015}, and \jayhorn~\cite{Kahsai2016} as well as efficient constraint solving tools such as \spacer~\cite{Komuravelli2014}, \eldarica~\cite{Hojjat2018}, and \hoice~\cite{Champion2018}.

In this paper we show that the same phenomenon---separating constraint generation from solving---can empower a wider class of verification problems.
To this end, we generalize \CHCS{} and introduce a new class of predicate Constraint Satisfaction Problems called \PCSPWFFN{} where constraints are arbitrary (i.e., possibly non-Horn) clauses modulo first-order theories over function variables and (possibly well-founded) predicate variables. 
We then show that \PCSPWFFN{} can encode a wider range of verification problems including but not limited to termination/non-termination verification and even linear-time \& branching-time temporal verification (full modal mu-calculus model checking) of programs written in various languages.
All these become possible due to the increased expressiveness: we show that \PCSPWFFN{} can express queries for a new extension of \CLP{}, denoted \MuCLP{}, that has arbitrarily nested inductive and co-inductive predicates. \MuCLP{} can naturally encode the above classes of verification problems and subsumes first-order fixpoint logics (which have recently been applied to temporal verification of imperative and functional programs~\cite{Kobayashi2019,Nanjo2018}).

The first part of this paper is  a sound and complete reduction algorithm from \MuCLP{} to \PCSPWFFN{}. The algorithm generalizes the recently proposed deductive system for the validity of first-order fixpoint logic~\cite{Nanjo2018} to \MuCLP{}.  It obtains a collection $\clauses$ of clause constraints that have placeholder function variables $\fnp{T}, \fnp{U},\ldots$ and predicate variables $P,Q,\ldots$, including some for well-founded relations $\wfp{R}, \wfp{S},\ldots$ (see Section~\ref{sec:pcsp} for the definition).

Next, we give a constraint solving method for \PCSPWFFN{} based on \emph{stratified} CounterExample-Guided Inductive Synthesis (CEGIS) of (co-)inductive invariants, ranking functions, and Skolem functions witnessing existential quantifiers.  Stratified CEGIS combines CEGIS~\cite{Solar-Lezama2006} with stratified families of templates~\cite{Jhala2006,Terauchi2015} to achieve relative completeness, a theoretical guarantee of convergence, and a faster and stable convergence by avoiding the overfitting problem of expressive templates to counterexamples~\cite{Padhi2019}.
The constraint solving method naturally generalizes a number of previous techniques developed for \CHCS{} solving and invariant/ranking function synthesis to the new class \PCSPWFFN{}.  It proceeds with an iterative algorithm that attempts to discover appropriate functions/predicates or counterexamples to the given \PCSPWFFN{} $\clauses$.  Each iteration consists of a synthesis phase that attempts to guess the function/predicate variables (represented as a function/predicate substitution over the stratified families of templates) and a validation phase that determines whether the guess was valid.  The validation is done by substituting the guess to $\clauses$ and using an SMT solver to determine whether $\clauses$ is satisfiable.  
When the substituion yields a satisfiable $\clauses$ we conclude the \MuCLP{} queries to be valid. Meanwhile, iterations maintain example instances of $\clauses$ from failed attempts by previous candidates and if these examples become unsatisfiable, we conclude the queries to be invalid.

We have implemented the above framework.  The implementation supports various widely used background theories: Booleans, linear integer and rational arithmetic.  We have applied our tool to a diverse collection of verification problems (modal mu-calculus, CTL*, CTL, LTL, termination, and safety) and obtained promising results.  The benchmark problems used for experiments go beyond the capabilities of the existing related tools (such as \CHCS{} solvers and program verification tools).

The rest of the paper is organized as follows.  Section~\ref{sec:overview} gives a brief overview with some examples.  Section~\ref{sec:muclp} defines \MuCLP{} and discusses its expressiveness and applications.  Section~\ref{sec:pcsp} defines \PCSPWFFN{} and Section~\ref{sec:reduct} formalizes the reduction from \MuCLP.  We present our constraint solving method for \PCSPWFFN{} based on stratified CEGIS in Section~\ref{sec:csolve}.  Section~\ref{sec:eval} reports on the implementation and experimental evaluation of the presented framework.  We discuss related work in Section~\ref{sec:related} and conclude with a remark on future work in Section~\ref{sec:conc}.

%% file: overview.tex
We now highlight the contributions of our work through a series of representative examples that we will return to later in the paper.

\subsection{Modeling language \MuCLP{}: Generalizing CLP.}
\input{overview_muclp}

\subsection{Intermediate representation \PCSPWFFN{}.}
\input{overview_pcsp}

\subsection{CounterExample-Guided Inductive Synthesis}
\input{overview_stratified_cegis}

%% file: overview_muclp.tex
Our first contribution is \MuCLP{}.  It generalizes CLP to allow describing a wider range of verification problems.
Let us consider the termination verification problem of the following program obtained from the benchmark set of the \function{} tool~\cite{Urban2013,Urban2014}, which is available from its web interface\footnote{\url{https://www.di.ens.fr/\~urban/FuncTion.html}}:
\begin{alltt}
while (x1 >= 0 && x2 >= 0) \{
  if (nondet()) \{ while (x2 <= 10 && nondet()) \{ x2 = x2 + 1; \}
                  x1 = x1 - 1; \}
  x2 = x2 - 1; \}
\end{alltt}
where \verb|nondet()| returns a non-deterministic Boolean value.  This program is \emph{always terminating} for any external integer inputs \verb|x1|,\verb|x2| and any internal Boolean non-deterministic choices.\footnote{Note that the termination is witnessed by, for example, the lexicographic order of \verb|x1|,\verb|x2|.}

The termination verification problem for the program can be \emph{modularly} encoded as the following \MuCLP{} $\mathcal{P}_{\mathrm{term}}$ using \emph{both least and greatest fixpoints} in the style of extended refinement type systems \cite{Unno2018,Nanjo2018}:
\begin{align*}
\begin{array}{rl}
 \textrm{Query}: & \forall x_1,x_2 \colon \INT.\ I(x_1,x_2) \\
 \textrm{Program}: &
 \left\{
\begin{array}{ll}
  I(x_1,x_2) &=_\mu
    \begin{array}{l}
    \neg (x_1 \geq 0 \land x_2 \geq 0)\ \lor \\
    \left(\begin{array}{l}
    J(x_2)\ \land \\
    (\forall x_2' \colon \INT.\ \mathit{NP}(x_2,x_2') \lor I(x_1-1,x_2'-1))\ \land \\
    I(x_1,x_2-1)
    \end{array}\right)
    \end{array} \\
  J(x_2) &=_\mu \neg (x_2 \leq 10) \lor J(x_2+1) \\
  \mathit{NP}(x_2,x_2') &=_\nu
  \neg\left(\begin{array}{l}
  \neg (x_2 \leq 10) \land x_2'=x_2\ \lor \\
  x_2 \leq 10 \land (\neg \mathit{NP}(x_2+1,x_2') \lor x_2'=x_2)
  \end{array}\right)
\end{array}
\right.
\end{array}
\end{align*}
Here, $J$ is an \emph{inductive predicate} defined as the \emph{least fixpoint} of the function $\mathcal{F}(J)=\lambda x_2.\ \neg(x_2 \leq 10) \lor J(x_2+1)$ over predicates (indicated by $\mu$ in $=_\mu$).  Likewise, $I$ is also an inductive predicate and is defined as a least fixpoint.  By contrast, $\mathit{NP}$ is a \emph{co-inductive predicate} defined as the \emph{greatest fixpoint} of the function $\mathcal{G}(\mathit{NP})=\lambda (x_2,x_2').\ \neg(\neg(x_2 \leq 10) \land x_2'=x_2 \lor x_2 \leq 10 \land (\neg \mathit{NP}(x_2+1,x_2') \lor x_2'=x_2))$.
Intuitively, $I(x_1,x_2)$ and $J(x_2)$ characterize the weakest pre-conditions for the termination of the outer and the inner loops, respectively; Note that the inner loop always terminates (regardless of non-deterministic choices) if and only if $\neg(x_2 \leq 10)$ eventually holds after a \emph{finite} number of iterations of incrementing $x_2$, which is here enforced by the \emph{least-fixpoint} definition of $J$.
$\mathit{NP}(x_2,x_2')$ denotes the \emph{complement} of the following inductive predicate $\mathit{P}(x_2,x_2')$ which characterizes the strongest post-condition of the inner loop:
\begin{align*}
  \mathit{P}(x_2,x_2') &=_\mu
  \neg (x_2 \leq 10) \land x_2'=x_2 \lor
  x_2 \leq 10 \land (\mathit{P}(x_2+1,x_2') \lor x_2'=x_2).
\end{align*}
In the definition of $I$, $\mathit{NP}(x_2,x_2')$ is used to bind $x_2'$ to a possible value of the program variable \verb|x2| upon the termination of the inner loop, encapsulating the internal behavior of the inner loop.
Thus, the query formula is valid if and only if the program is always terminating for all initial integer valuations of \verb|x1| and \verb|x2| and for all internal non-deterministic choices.  Though we could encode the termination verification problem using only least fixpoints by regarding the given program as a single monolithic transition system, this example demonstrates an advantage of the use of \MuCLP{} for \emph{modularly and naturally} encoding verification problems.

To demonstrate another advantage of \MuCLP{}, let us now consider verifying non-termination for the same program.  Thanks to the expressiveness of \MuCLP{}, this can be encoded as the following \MuCLP{} $\mathcal{P}_{\mathrm{nterm}}$, which is simply the De Morgan dual of $\mathcal{P}_{\mathrm{term}}$:
\begin{align*}
\begin{array}{rl}
 \textrm{Query}: & \exists x_1,x_2 \colon \INT.\ \mathit{NI}(x_1,x_2) \\
 \textrm{Program}: &
 \left\{
\begin{array}{ll}
  \mathit{NI}(x_1,x_2) &=_\nu
    \begin{array}{l}
    x_1 \geq 0 \land x_2 \geq 0\ \land \\
    \left(\begin{array}{l}
    \mathit{NJ}(x_2)\ \lor \\
    (\exists x_2' \colon \INT.\ \mathit{P}(x_2,x_2') \land \mathit{NI}(x_1-1,x_2'-1))\ \lor \\
    \mathit{NI}(x_1,x_2-1)
    \end{array}\right)
    \end{array} \\
  \mathit{NJ}(x_2) &=_\nu x_2 \leq 10 \land \mathit{NJ}(x_2+1) \\
  \mathit{P}(x_2,x_2') &=_\mu
  \neg (x_2 \leq 10) \land x_2'=x_2 \lor
  x_2 \leq 10 \land (\mathit{P}(x_2+1,x_2') \lor x_2'=x_2)
\end{array}
\right.
\end{array}
\end{align*}
Intuitively, $\mathit{NI}(x_1,x_2)$ and $\mathit{NJ}(x_2)$ respectively characterize the weakest pre-conditions for the \emph{non-termination} of the outer and the inner loops, which generalize the recurrent sets~\cite{Gupta2008} whose inhabitant witnesses the non-termination of the given program, for the purpose of modular encoding.  Recall that $\mathit{P}(x_2,x_2')$ characterizes the strongest post-condition of the inner loop.

Our validity checker \muval{} for \MuCLP{} tries to solve the primary $\mathcal{P}_{\mathrm{term}}$ and the dual $\mathcal{P}_{\mathrm{nterm}}$ in parallel.  The primal-dual approach turns out to be particularly useful for branching-time temporal verification where we found either the primary or the dual is often easier to solve than the other (cf.~Section~\ref{sec:eval}).

As we show in Section~\ref{sec:apps_temp} and Appendix~\ref{sec:apps}, \MuCLP{} is expressive enough to naturally encode a diverse class of program verification problems. 
 \begin{itemize}
 \item Linear-time temporal verification of labeled transition systems.  Section~\ref{sec:apps_temp} explains a reduction from ($\omega$-)regular model checking where the specifications are given as B\"{u}chi word automata (which strictly subsume LTL).
\item Bisimulation and bisimilarity verification between labeled transition systems (Appendix~\ref{sec:apps_coind}).
\item Infinite state, infinite duration games.  Safety games, reachability games and so-called LTL games (Appendix~\ref{sec:apps_game}).
\end{itemize}
Also, it immediately follows from existing results that \MuCLP{} can encode:
\begin{itemize}
\item Linear-time temporal verification of functional programs~\cite{Nanjo2018,Kobayashi2019}.
\item Branching-time temporal verification.  A reduction algorithm from modal mu-calculus model checking of imperative programs is shown in \cite{Kobayashi2019}.
\end{itemize}

%% file: overview_pcsp.tex
\MuCLP{} is a very expressive language, and existing verification intermediate representations such as Constrained Horn Clauses (\CHCS{})~\cite{Bjorner2015a} are not powerful enough to capture the full class of \MuCLP{}.  We therefore introduce a new verification intermediate representation: a class of \emph{predicate} Constraint Satisfaction Problems denoted \PCSPWFFN{}.  \PCSPWFFN{} is a generalization of \CHCS{} to arbitrary clauses, function variables, and well-foundedness constraints over predicate variables.

We will present a sound and complete reduction from \MuCLP{} validity to \PCSPWFFN{} satisfiability in Section~\ref{sec:reduct}.  It is inspired by a recently proposed deductive system for first-order fixpoint logic~\cite{Nanjo2018} that eliminates least and greatest fixpoints by over- and under-approximations via (co-)inductive invariants and well-founded relations, and eliminates quantifiers by Skolemization.
For the termination verification problem $\mathcal{P}_{\mathrm{term}}$, our reduction gives the following set of clauses whose term variables are implicitly universally quantified:
\begin{align*}
\left\{
\begin{array}{ll}
\uapprox{I}(x_1,x_2), & (1) \\
\uapprox{I}(x_1,x_2) \imply
    \begin{array}{l}
    \neg (x_1 \geq 0 \land x_2 \geq 0)\ \lor \\
    \left(\begin{array}{l}
    \uapprox{J}(x_2) \land (\uapprox{\mathit{NP}}(x_2,x_2') \lor \uapprox{I}(x_1-1,x_2'-1) \land \wfp{I}(x_1,x_2,x_1-1,x_2'-1))\ \land \\
    \uapprox{I}(x_1,x_2-1) \land \wfp{I}(x_1,x_2,x_1,x_2-1)
    \end{array}\right)
    \end{array}, & (2) \\
  \uapprox{J}(x_2) \imply \neg(x_2 \leq 10) \lor \uapprox{J}(x_2+1) \land \wfp{J}(x_2,x_2+1), & (3) \\
  \uapprox{\mathit{NP}}(x_2,x_2') \imply
  \neg\left(\begin{array}{l}
  \neg (x_2 \leq 10) \land x_2'=x_2\ \lor \\
  x_2 \leq 10 \land (\neg\uapprox{\mathit{NP}}(x_2+1,x_2') \lor x_2'\neq x_2)
  \end{array}\right) & (4)
\end{array}
\right\}
\end{align*}
Here $\uapprox{I}$, $\uapprox{J}$, and $\uapprox{\mathit{NP}}$ are predicate variables that represent an under-approximation of the (co-)inductive predicates $I$, $J$, and $\mathit{NP}$, respectively.  $\wfp{I}$ and $\wfp{J}$ are \emph{well-founded} predicate variables that are required to represent a well-founded relation and used here to enforce a \emph{bounded}
unfolding of the inductive predicates $I$ and $J$, respectively; Note here that, in the third clause, $\wfp{J}(x_2,x_2+1)$ requires that the formal argument $x_2$ of $J$ and the actual argument $x_2+1$ of the recursive call to $J$ are related by a well-founded relation.  Similarly, $\wfp{I}(x_1,x_2,x_1-1,x_2'-1)$ and $\wfp{I}(x_1,x_2,x_1,x_2-1)$ in the second clause require that the pair $(x_1,x_2)$ of the formal arguments and the pairs of the actual arguments of the two recursive calls are respectively related by a well-founded relation.  This transformation of inductive predicates generalizes \emph{binary reachability analysis} that has been studied for termination verification of imperative~\cite{Podelski2004a,Cook2006} and functional programs~\cite{Kuwahara2014}.
The obtained \PCSPWFFN{} is \emph{satisfiable}: our satisfiability checker \pcsat{} for \PCSPWFFN{} reports the following satisfying predicate assignment:
\begin{align*}
\uapprox{I}(x_1,x_2) &\mapsto \top,\quad \uapprox{J}(x_2) \mapsto x_2 \geq 0,\quad \uapprox{\mathit{NP}}(x_1,x_2) \mapsto \bot \\
\wfp{I}(x_1,x_2,x_1',x_2') &\mapsto x_1 \geq 0 \land x_1+x_2 \geq 0 \land (x_1 > x_1' \lor x_1 \geq x_1' \land x_1+x_2 > x_1'+x_2') \\
\wfp{J}(x_2,x_2') &\mapsto x_2 \geq 0 \land x_2' \geq 0 \land  \max(22-x_2,x_2) \geq 0 \land \max(22-x_2,x_2) > \max(22-x_2',x_2')
\end{align*}
Here, $\max(t_1,t_2)$ represents the maximum of integer terms $t_1$ and $t_2$, $\wfp{I}$ and $\wfp{J}$ represent the well-founded relations respectively induced by the lexicographic ranking function $\lambda (x_1,x_2). (x_1,x_1+x_2)$ and the piecewise-defined ranking function $\lambda x_2 \geq 0. \max(22-x_2,x_2)$.

For the non-termination verification problem $\mathcal{P}_{\mathrm{nterm}}$, our reduction gives the following \PCSPWFFN{}:
\begin{align*}
\left\{
\begin{array}{ll}
\uapprox{\mathit{NI}}(S_{\lambda},T_{\lambda}), & (1) \\
  \uapprox{\mathit{NI}}(x_1,x_2) \imply
    \begin{array}{l}
    x_1 \geq 0 \land x_2 \geq 0\ \land \\
    \left(\begin{array}{l}
    \uapprox{\mathit{NJ}}(x_2)\ \lor \\
    (\uapprox{P}(x_2,U_{\lambda}(x_1,x_2)) \land \uapprox{\mathit{NI}}(x_1-1,U_{\lambda}(x_1,x_2)-1))\ \lor \\
    \uapprox{\mathit{NI}}(x_1,x_2-1)
    \end{array}\right)
    \end{array}, & (2) \\
  \uapprox{\mathit{NJ}}(x_2) \imply x_2 \leq 10 \land \uapprox{\mathit{NJ}}(x_2+1), & (3) \\
  \uapprox{P}(x_2,x_2') \imply
  \left(\begin{array}{l}
  \neg (x_2 \leq 10) \land x_2'=x_2\ \lor \\
  x_2 \leq 10 \land (\uapprox{P}(x_2+1,x_2') \land \wfp{P}(x_2,x_2',x_2+1,x_2') \lor x_2'=x_2)
  \end{array}\right) & (4)
\end{array}
\right\}
\end{align*}
Here $\uapprox{\mathit{NI}}$, $\uapprox{\mathit{NJ}}$, and $\uapprox{P}$ are predicate variables that represent an under-approximation of the (co-)inductive predicates $\mathit{NI}$, $\mathit{NJ}$, and $P$, respectively.  $\wfp{P}$ is a well-founded predicate variable used here to enforce a bounded unfolding of $P$.
$S_{\lambda}$, $T_{\lambda}$, and $U_{\lambda}$ are \emph{function} variables that represent total functions to be synthesized and used here to Skolemize the existential quantification of the term variables $x_1,x_2$ in the query and $x_2'$ in the body of $\mathit{NJ}$, respectively.\footnote{Note that we regard $S_{\lambda}$, $T_{\lambda}$ as integer variables that represent integer functions of the arity 0.}
Not surprisingly, this \PCSPWFFN{} is \emph{unsatisfiable}.

%% file: overview_stratified_cegis.tex
Our reduction from \MuCLP{} to \PCSPWFFN{} may generate constraints that go beyond the class of \CHCS{}.  We thus present a new constraint solving method that can handle the general class of constraints, which is formally defined in Section~\ref{sec:pcsp} and here explained informally using the following example \PCSPWFFN{}:
\[ \clauses \defeq
\left\{
\begin{array}{ll}
n \geq 0 \imply \uapprox{X}(n), & (1) \\
\uapprox{X}(x) \imply (\uapprox{Y}(x) \land \uapprox{X}(x+1)), & (2) \\
\uapprox{Y}(y) \imply (y=0 \lor \uapprox{Y}(y-1) \land \wfp{Y}(y, y-1))  \qquad & (3)
\end{array}
\right\}
\]
\input{example_scegis}

%% file: example_scegis.tex
Our constraint solving method is based on a CounterExample Guided Inductive Synthesis (CEGIS) approach to finding a solution of the given constraint set $\clauses$, i.e., a predicate substitution for $\uapprox{X},\uapprox{Y},\wfp{Y}$ that satisfies all the three formulas in $\clauses$ and the well-foundedness condition of $\wfp{Y}$.  Our method is designed as a general constraint solving schema and this paper presents an instantiation of the schema based on template-based synthesis~\cite{Sharma2013a,Garg2014}.  This section is mostly dedicated to informally reviewing well-known CEGIS with template-based synthesis in order to make the paper self-contained.  Detailed exposition, in particular, our extensions with stratified families of function/predicate templates and unsat-core-based template refinement, are given in Section~\ref{sec:csolve}).

Our method first prepares predicate templates $\template_{\uapprox{X}}$, $\template_{\uapprox{Y}}$, $\template_{\wfp{Y}}$, with unknown parameters to be inferred, respectively for the predicate variables $\uapprox{X}$, $\uapprox{Y}$, $\wfp{Y}$ to restrict the solution space to be explored.
For example, let us here use the templates:\footnote{Our stratified template families further support templates of more general shapes: disjunctions of conjunctions of atomic formulas for ordinary predicate variables, well-founded relation templates induced by lexicographic piecewise-defined affine ranking for well-founded predicate variables, and piecewise-defined affine function templates for function variables.}
\begin{align*}
\template_{\uapprox{X}} &\defeq \lambda x. a \cdot x + b \geq 0, \qquad
\template_{\uapprox{Y}} \defeq \lambda y. c \cdot y + d \geq 0, \\
\template_{\wfp{Y}} &\defeq \lambda (z,z'). d \cdot z + e \geq 0 \land d \cdot z + e > d \cdot z' + e.
\end{align*}
Here, $a,b,c,d,e$ are unknown parameters of the predicate templates.  Note that the form of the predicate template $\template_{\wfp{Y}}$ for $\wfp{Y}$ guarantees that $\wfp{Y}$ is a well-founded relation for any valuation of $d,e$: the function $\lambda z.\ d \cdot z + e$ represents an affine ranking function whose return value strictly decreases, when the input changes from $z$ to $z'$. These templates are geared to the background theory, but other templates could be used for other theories.

Our constraint solving schema then iteratively accumulates examples $\examples$ of the constraints in $\clauses$ by instantiating term variables to concrete values in a counterexample guided manner and enumerates candidate solutions using $\examples$ until a genuine solution for $\clauses$ is obtained.  More specifically, at each iteration $i$, our schema consists of two phases: {\bf Synthesis Phase} asks a synthesizer to obtain a candidate solution $\sigma^{(i)}$ that satisfies the set of examples $\examples^{(i)}$ and {\bf Validation Phase} checks whether $\sigma^{(i)}$ is a genuine solution of $\clauses$.  If it is the case, our schema returns $\sigma^{(i)}$ as a solution and otherwise repeats with $\examples^{(i+1)}$ obtained from $\examples^{(i)}$ by adding new examples that are not satisfied by $\sigma^{(i)}$.

We now illustrate this procedure, using the running example. There will be four iterations, each with two phases. 

\paragraph{First iteration.} In the first iteration, we have no examples of the constraints yet, so we start with examples $\examples^{(1)}=\emptyset$. The template-based synthesizer, using {\it e.g.} an SMT solver, may generate any candidate solution such as:
\begin{align*}
\sigma^{(1)} &\defeq \theta^{(1)}(\{\uapprox{X} \mapsto \template_{\uapprox{X}}, \uapprox{Y} \mapsto \template_{\uapprox{Y}}, \wfp{Y} \mapsto \template_{\wfp{Y}}\}), \\
\theta^{(1)} &\defeq \{a \mapsto 0, b \mapsto 0, c \mapsto 0, d \mapsto 0, e \mapsto 0\}
\end{align*}
where $\theta^{(1)}$ is a substitution of values for
template parameters. 
We next use this parameter assignment, substituting it back into the templates. In this example, substituting $a \mapsto 0$ and $b\mapsto 0$ into $\template_{\uapprox{X}}$, yields $\lambda x.0+0\geq 0$, or $\lambda x.\top$.
We similarly obtain $\template_{\uapprox{Y}} = \lambda y.\top$ but obtain $\template_{\wfp{Y}}=\lambda (z,z').\bot$. Together, we have
\begin{align*}
\sigma^{(1)} \equiv \{\uapprox{X} \mapsto \lambda x. \top, \uapprox{Y} \mapsto \lambda y. \top, \wfp{Y} \mapsto \lambda (z,z'). \bot\}.
\end{align*}

We now enter the second phase of the first iteration: we need to check whether $\sigma^{(1)}$ is a genuine solution of $\clauses$. We substitute $\sigma^{(1)}$ back into $\clauses$ and for Eqn.~3, to obtain
\begin{align*}
\sigma^{(1)}(\uapprox{Y})(y) \Rightarrow (y=0 \lor \sigma^{(1)}(\uapprox{Y})(y-1) \land \sigma^{(1)}(\wfp{Y})(y, y-1)).
\end{align*}
Using an SMT solver, we can find that this is not valid and, thus, the constraint is not satisfied. An SMT solver may, for example, generate a model $y \mapsto 1$, which gives us an 
example $\uapprox{Y}(1) \Rightarrow \uapprox{Y}(0) \land \wfp{Y}(1, 0)$ of
$\clauses$ that is not satisfied by $\sigma^{(1)}$.
This provides us with a new \emph{example} $\examples^{(2)}$ of $\clauses$ that is not satisfied by $\sigma^{(1)}$:
\begin{align*}
\examples^{(2)} \defeq \{\uapprox{Y}(1) \Rightarrow \uapprox{Y}(0) \land \wfp{Y}(1, 0)\}.
\end{align*}

\paragraph{Remaining iterations 2, 3, and 4.} The next iterations proceed similarly, and yield the following solutions and examples:
\begin{align*}
\sigma^{(2)} &\equiv \{\uapprox{X} \mapsto \lambda x. \top, \uapprox{Y} \mapsto \lambda y. \top, \wfp{Y} \mapsto \lambda (z,z'). z \geq 0 \land z > z'\} \\
\examples^{(3)}&=\examples^{(2)} \cup \{\uapprox{Y}(-1) \Rightarrow \uapprox{Y}(-2) \land \wfp{Y}(-1, -2)\} \\
\sigma^{(3)} &\equiv \{\uapprox{X} \mapsto \lambda x. \top, \uapprox{Y} \mapsto \lambda y. y \geq 0, \wfp{Y} \mapsto \lambda (z,z'). z \geq 0 \land z > z'\}\\
\examples^{(4)}&=\examples^{(3)} \cup \{\uapprox{X}(-1) \Rightarrow (\uapprox{Y}(-1) \land \uapprox{X}(0))\}\\
\sigma^{(4)} &\equiv \{\uapprox{X} \mapsto \lambda x. x \geq 0, \uapprox{Y} \mapsto \lambda y. y \geq 0, \wfp{Y} \mapsto \lambda (z,z'). z \geq 0 \land z > z'\}
\end{align*}
From iteration 3, we have refined $\uapprox{Y}$, requiring that it constrains $y$ to be positive. This constraint eliminates the issue that arose from parameters to $\wfp{Y}$ being negative. 
Iteration 4 similarly teaches us that $x$ must be positive. 
In the final iteration's second phase, we find that
$\sigma^{(4)}$ is a genuine solution of $\clauses$ and exit the procedure.

%% file: muclp.tex
This section defines the syntax and the semantics of the extension \MuCLP{} of constraint logic programs \CLP{}~\cite{Jaffar1994} with arbitrarily nested inductive and co-inductive predicates.  We also discuss its application to temporal verification.

\subsection{Syntax}
\label{sec:syntax}
\input{syntax}

\subsection{Semantics}
\label{sec:semantics}
\input{semantics}

\subsection{Application to Temporal Property Verification}
\label{sec:apps_temp}
\input{apps_temp}

%% file: syntax.tex
Let $\theory$ be a (possibly many-sorted) first-order theory with the signature $\Sigma$.  The syntax of $\theory$-formulas and $\theory$-terms is:
\begin{align*}
  \text{(formulas)} \;
  \phi &::= X(t_1,\dots, t_{\arity{X}}) \mid p(t_1, \dots, t_{\arity{p}})
  \mid \neg \phi
  \mid \phi_1 \lor \phi_2
  \mid \phi_1 \land \phi_2
  \mid \exiC{x}{s}{\phi} \mid \allC{x}{s}{\phi} \\
  \text{(terms)} \;
  t &::= x \mid F(t_1, \dots, t_{\arity{F}}) \mid f(t_1, \dots, t_{\arity{f}})
\end{align*}
Here, the meta-variables $x$, $X$, and $F$ respectively range over 
term, predicate, and function variables.  The meta-variables $p$ and $f$ respectively denote predicate and function symbols of the signature $\Sigma$.  We use $s$ as a meta-variable ranging over sorts of the signature $\Sigma$.  We write $\propos$ for the sort of propositions and $s_1 \to s_2$ for the sort of functions from $s_1$ to $s_2$.  We henceforth regard a predicate variable as a function variable whose return sort is $\propos$.  We write $\arity{o}$ and $\sort{o}$ respectively for the arity and the sort of a syntactic element $o$.  A function $f$ represents a constant if $\arity{f}=0$.  We write $\ftv{\phi}$, $\fpv{\phi}$, and $\ffv{\phi}$ respectively for the set of free term, predicate, and function variables that occur in $\phi$.  Note that $\fpv{\phi} \subseteq \ffv{\phi}$ always holds.  We write $\myseq{x}$ for a sequence of term variables, $\length{\myseq{x}}$ for the length of $\myseq{x}$, and $\epsilon$ for the empty sequence.  We often abbreviate $\lnot \phi_1 \lor \phi_2$ as $\phi_1 \imply \phi_2$.  We henceforth consider only well-sorted formulas and terms.

A \MuCLP{}~$\mathcal{P}$ over the theory~$\theory$ is a sequence of mutually (co-)recursive equations of the form:\footnote{If we fix $\theory$ to integer arithmetic, \MuCLP{} coincides \muarith{}, a fixpoint logic with integer arithmetic studied in \cite{Lubarsky1993,Bradfield1999} and reformalized as hierarchical equation systems (\HES) in \cite{Kobayashi2019}.}
\begin{align*}
(X_1(\myseq{x}_1) =_{\alpha_1} \phi_1); \dots; (X_{m}(\myseq{x}_{m})=_{\alpha_m}\phi_m)
\end{align*}
Here, $\alpha_i\in\finset{\mu,\nu}$ and for any $i,j \in \finset{1,\dots,m}$,
$X_i$ may occur only positively in $\phi_j$.  An equation $X(\myseq{x}) =_\mu \phi$ that satisfies $\ffv{\phi} \subseteq \finset{X}$ represents the inductive predicate $\lfp{\myseq{x}}{\phi}$ defined as the least fixpoint of the function $\mathcal{F}(X)=\lambda \myseq{x}.\phi$ over predicates.  Similarly, $X(\myseq{x}) =_\nu \phi$ that satisfies $\ffv{\phi} \subseteq \finset{X}$ represents the co-inductive predicate $\gfp{\myseq{x}}{\phi}$ defined as the greatest fixpoint of $\mathcal{F}$.  Note here that $\mathcal{F}$ is monotonic because the bound predicate variable $X$ occurs only positively in the body $\phi$.  In cases where $\ffv{\phi} \subseteq \finset{X}$ does not hold, more sophisticated semantic treatment is required.  We formalize this point later in Section~\ref{sec:semantics}.
We define $\dom{\mathcal{P}}=\finset{X_1,\dots,X_m}$.  A \emph{query} for a \MuCLP{} $\mathcal{P}$ is defined as a $\theory$-formula $\phi$.
The De Morgan dual $\neg \mathcal{P}$ of a \MuCLP{} $\mathcal{P} = (X_i(\myseq{x}_i)=_\alpha \phi_i)_{i=1}^m$ is defined by $(X_i^\neg(\myseq{x}_i)=_{\neg \alpha} \sigma(\neg \phi_i))_{i=1}^m$ where
$\sigma \defeq \{X_1 \mapsto \neg X_1^\neg,\dots X_m \mapsto \neg X_m^\neg\}$, $\neg\mu \defeq \nu$, and $\neg\nu \defeq \mu$.

\begin{remark}
\label{rem:qelim}
Note that quantifiers over recursively enumerable (r.e.) domains (e.g., integers) can be eliminated in \MuCLP{}; We can encode $\exiC{x}{\INT}{\phi}$ and $\allC{x}{\INT}{\phi}$ with the bound integer variable $x$ respectively as $E(0)$ and $A(0)$ using the following inductive and co-inductive predicates $E$ and $A$:
\begin{align*}
  E(x) =_\mu \phi \lor [-x/x]\phi \lor E(x+1) \qquad
  A(x) =_\nu \phi \land [-x/x]\phi \land A(x+1)
\end{align*}
Intuitively, $E$ and $A$ are required to hold for some and for all integer $x$, respectively.  This encoding strategy, however, cannot apply to non r.e. domains like real numbers and is not useful in practice even for r.e. domains like rational numbers that have no simple way to enumerate all its elements.
This is the reason why we apply Skolemization via function variables instead in our reduction algorithm to \PCSPWFFN{} (see Section~\ref{sec:reduct} for details).
\qed
\end{remark}

%% file: semantics.tex
This section formalizes the denotational semantics of \MuCLP{}.  Let $\structure=(\fpdom, \Sigma, I)$ be the structure of the background first-order theory $\theory$.  Here, $\fpdom$ is the universe, $\Sigma$ is the signature, and $I$ is the interpretation function for the predicate and function symbols in $\Sigma$.  We write $\fpdom_s$ for the set of values in $\fpdom$ of the sort $s$.  In particular, We define $\fpdom_{\propos} \defeq \{\top, \bot\}$ for the sort $\propos$ of propositions.
For a sequence $\myseq{s}=s_1,\dots,s_m$ of sorts with $m \geq 0$, we write $\fpdom_{\myseq{s}}$ for the sequence $\fpdom_{s_1},\dots,\fpdom_{s_m}$.  We define $\fpdom_{(\myseq{s} \to s)} \defeq \fpdom_{\myseq{s}} \to \fpdom_{s}$.  We assume that $I(p) \in \fpdom_{\myseq{s}} \to \fpdom_\propos$ if $\sort{p}=\myseq{s} \to \propos$, and $I(f) \in \fpdom_{\myseq{s}} \to \fpdom_s$ if $\sort{f}=\myseq{s} \to s$.
We introduce the partially ordered sets $(\fpdom_{(\myseq{s} \to \propos)}, \sqsubseteq_{(\myseq{s} \to \propos)})$ by defining
\begin{align*}
    \sqsubseteq_\propos &\defeq \{(\top, \top), (\bot, \top), (\bot, \bot)\}, &
    \sqsubseteq_{(\myseq{s} \to \propos)} &\defeq \{(f, g) \mid
        \all
            {\myseq{v} \in \fpdom_{\myseq{s}}}
            {f(\myseq{v}) \sqsubseteq_\propos G(\myseq{v})}
    \}.
\end{align*}
The least upper bound $\sqcup_{(\myseq{s} \to \propos)}$ and the greatest lower bound $ \sqcap_{(\myseq{s} \to \propos)}$ operators with respect to $\sqsubseteq_{(\myseq{s} \to \propos)}$ are then defined as follows:
\begin{align*}
&\begin{array}{rclcrclcrclcrcl}
    \top \sqcap_{\propos} \top &\defeq& \top &&
    \top \sqcap_{\propos} \bot &\defeq& \bot &&
    \bot \sqcap_{\propos} \top &\defeq& \bot &&
    \bot \sqcap_{\propos} \bot &\defeq& \bot \\
    \top \sqcup_{\propos} \top &\defeq& \top &&
    \top \sqcup_{\propos} \bot &\defeq& \top &&
    \bot \sqcup_{\propos} \top &\defeq& \top &&
    \bot \sqcup_{\propos} \bot &\defeq& \bot
\end{array} \\
&\begin{array}{rclcrcl}
    f \sqcap_{(\myseq{s} \to \propos)} g &\defeq& \lamI{\myseq{v}}{\fpdom_{\myseq{s}}}{f(\myseq{v}) \sqcap_{\propos} g(\myseq{v})} &&
    f \sqcup_{(\myseq{s} \to \propos)} g &\defeq& \lamI{\myseq{v}}{\fpdom_{\myseq{s}}}{f(\myseq{v}) \sqcup_{\propos} g(\myseq{v})}
\end{array}
\end{align*}
Note that $(\fpdom_{\myseq{s} \to \propos}, \sqsubseteq_{\myseq{s} \to \propos})$ forms a complete lattice.  The least and greatest elements of $\fpdom_{\myseq{s} \to \propos}$ are $\lam{\myseq{x}}{\bot}$ and $\lam{\myseq{x}}{\top}$ respectively.

Given a $\theory$-formula $\phi$ and an interpretation $\rho$ of free term and function/predicate variables in $\phi$, we write $\denote{}{\phi}(\rho)$ for the truth value of $\phi$ which is defined as follows: 
\begin{align*}
&\begin{array}{rclcrcl}
\denote{}{\allC{x}{s}{\phi}}(\rho) &\defeq& {\textstyle \bigsqcap}_{\propos} \myset{\denote{}{\phi}(\rho\finset{x\mapsto v})}{v \in \fpdom_s} &&
\denote{}{X(\myseq{t})}(\rho) &\defeq& \rho(X)(\denote{}{\myseq{t}}(\rho)) \\
\denote{}{\exiC{x}{s}{\phi}}(\rho) &\defeq& {\textstyle \bigsqcup}_{\propos} \myset{\denote{}{\phi}(\rho\finset{x\mapsto v})}{v \in \fpdom_s} &&
\denote{}{p(\myseq{t})}(\rho) &\defeq& I(p)(\denote{}{\myseq{t}}(\rho)) \\
\multirow{2}{*}{$\denote{}{\lnot\phi}(\rho)$} &\multirow{2}{*}{$\defeq$} &
\multirow{2}{*}{
$\begin{cases}
\top & (\denote{}{\phi}(\rho)=\bot) \\
\bot & (\denote{}{\phi}(\rho)=\top)
\end{cases}$
} &&
\denote{}{\phi_1 \land \phi_2}(\rho) &\defeq& \denote{}{\phi_1}(\rho) \sqcap_{\propos} \denote{}{\phi_2}(\rho) \\
&&&& \denote{}{\phi_1 \lor \phi_2}(\rho) &\defeq& \denote{}{\phi_1}(\rho) \sqcup_{\propos} \denote{}{\phi_2}(\rho)
\end{array} \\
&\begin{array}{rclcrclcrcl}
\denote{}{x}(\rho) &\defeq& \rho(x) &&
\denote{}{F(\myseq{t})}(\rho) &\defeq& \rho(F)(\denote{}{\myseq{t}}(\rho))
&&
\denote{}{f(\myseq{t})}(\rho) &\defeq& I(f)(\denote{}{\myseq{t}}(\rho)) \end{array}
\end{align*}
Here, we assume that $\rho(x) \in \fpdom_{\sort{x}}$, $\rho(X) \in \fpdom_{\myseq{s}} \to \fpdom_\propos$ if $\sort{X}=\myseq{s} \to \propos$, and $\rho(F) \in \fpdom_{\myseq{s}} \to \fpdom_s$ if $\sort{X}=\myseq{s} \to s$.
We write $\rho \models \phi$ if and only if $\denote{}{\phi}(\rho')=\top$ holds for any extension $\rho'$ of $\rho$ for the term and function/predicate variables in $(\ftv{\phi} \cup \ffv{\phi}) \setminus \dom{\rho}$, where $\dom{\rho}$ represents the domain of $\rho$.  We say the given formula $\phi$ is \emph{valid} and write $\models \phi$ if and only if $\emptyset \models \phi$ holds.

Given a \MuCLP{} $\mathcal{P}$ and an interpretation $\rho$ of free term and function/predicate variables in $\mathcal{P}$, we write $\sem{\mathcal{P}}(\rho)$ for the predicate interpretation for $\dom{\mathcal{P}}$ induced by $\mathcal{P}$ which is defined by:
\begin{align*}
&\begin{array}{rl}
  \denote{}{\epsilon}(\rho) &\defeq \emptyset \\
  \denote{}{(X(\myseq{x}) =_{\alpha} \phi); \mathcal{P}}(\rho) &\defeq
  \denote{\mathcal{P}}{X(\myseq{x})=_{\alpha}\phi}(\rho)
    \uplus \sem{\mathcal{P}}(\rho\uplus\sem{X(\myseq{x}) =_{\alpha} \phi}_{\mathcal{P}}(\rho))
\end{array} \\
&\begin{array}{rl}
  \denote{\mathcal{P}}{X(\myseq{x}) =_{\alpha} \phi}(\rho) &\defeq
    \left\{X \mapsto \FP{\sort{X}}{\alpha}{
      \lambda q.
      \lambda \myseq{v}.
          \denote{}{\phi}
          \left(\rho\{X \mapsto q,\myseq{x}\mapsto\myseq{v}\} \uplus
           \denote{}{\mathcal{P}}\left(\rho\{X \mapsto q\}\right)\right)
      }\right\}
\end{array}
\end{align*}
where $\dom{\rho} \cap \dom{\mathcal{P}} = \emptyset$ and the fixpoint operator $\FP{\myseq{s} \to \propos}{\alpha}{\bullet}$ is defined by:
\begin{align*}
    \FP{\myseq{s} \to \propos}{\mu}{F} &\defeq {\textstyle \bigsqcap}_{(\myseq{s} \to \propos)}
        \myset{q \in \fpdom_{\myseq{s} \to \propos}}{F(q) \sqsubseteq_{(\myseq{s} \to \propos)} q} \\
    \FP{\myseq{s} \to \propos}{\nu}{F} &\defeq {\textstyle \bigsqcup}_{(\myseq{s} \to \propos)}
        \myset{q \in \fpdom_{\myseq{s} \to \propos}}{q \sqsubseteq_{(\myseq{s} \to \propos)} F(q)}
\end{align*}
We write $\mathcal{P} \models \phi$ if and only if $\denote{}{\mathcal{P}}(\emptyset) \models \phi$ holds.

\begin{example}
Let us consider \MuCLP{} $\mathcal{P}_{\nu\mu} \defeq (X =_{\nu} X \land Y);(Y =_{\mu} X \lor Y)$ and \MuCLP{} $\mathcal{P}_{\mu\nu} \defeq (Y =_{\mu} X \lor Y); (X =_{\nu} X \land Y)$.  Note that the semantics of $\mathcal{P}_{\nu\mu}$ and $\mathcal{P}_{\mu\nu}$ are different as shown below, though the definition of $\mathcal{P}_{\nu\mu}$ and $\mathcal{P}_{\mu\nu}$ only differ in the order of the equations:
\begin{align*}
  \denote{}{\mathcal{P}_{\nu\mu}}(\emptyset) &=
    \denote{(Y =_{\mu} X \lor Y)}{X=_{\nu} X \land Y}(\emptyset) \uplus
    \denote{\epsilon}{Y=_{\mu} X\lor Y}
      (\denote{(Y=_{\mu} X\lor Y)}{X=_{\nu}X\land Y}(\emptyset)) \\
    &= \rho^{\nu\mu} \uplus \denote{\epsilon}{Y=_{\mu} X \lor Y}(\rho^{\nu\mu}) \\
    &= \finset{X\mapsto \top} \uplus
       \denote{\epsilon}{Y=_{\mu} X \lor Y}(\finset{X\mapsto \top}) \\
    &= \left\{ X \mapsto \top, Y \mapsto \top \right\}
\end{align*}
where
\begin{align*}
    \rho^{\nu\mu}
    &= \left\{X \mapsto
      \FP{\propos}{\nu}{\lambda q.
      \denote{}{X\land Y}(\finset{X \mapsto q} \uplus \denote{}{Y =_{\mu} X \lor Y}(\finset{X \mapsto q}))
      }\right\} \\
    &= \left\{X \mapsto
      {\textstyle \bigsqcup}_{\propos}
      \setsc{q \in \fpdom_{\propos}}{
        q \sqsubseteq_{\propos}
        \denote{}{X\land Y}(\finset{X \mapsto q} \uplus \denote{}{Y =_{\mu} X \lor Y}(\finset{X \mapsto q}))
      }\right\} \\
    &= \left\{X \mapsto
      {\textstyle \bigsqcup}_{\propos}
      \setsc{q \in \fpdom_{\propos}}{
        q \sqsubseteq_{\propos}
        \denote{}{X\land Y}(\finset{X \mapsto q} \uplus \rho_q^{\mu})
      }\right\} \\
    &= \left\{X \mapsto
      {\textstyle \bigsqcup}_{\propos}
      \setsc{q \in \fpdom_{\propos}}{
        q \sqsubseteq_{\propos}
        \denote{}{X\land Y}(\finset{X \mapsto q,Y \mapsto q})
      }\right\} \\
    &= \left\{X \mapsto \top\right\} \\
    \rho_q^{\mu}
    &= \left\{Y \mapsto
    \FP{\propos}{\mu}{\lambda q'.\denote{}{X\lor Y}
                    (\finset{X \mapsto q,Y \mapsto q'})}
      \right\} \\
    &= \left\{Y \mapsto q\right\} \\
  \denote{}{\mathcal{P}_{\mu\nu}}(\emptyset) &=
    \denote{(X=_{\nu} X \land Y)}{Y =_{\mu} X \lor Y}(\emptyset) \uplus
    \denote{\epsilon}{X=_{\nu}X\land Y}
      (\denote{(X=_{\nu} X \land Y)}{Y=_{\mu} X\lor Y}(\emptyset)) \\
      &= \finset{Y \mapsto \bot} \uplus 
          \denote{\epsilon}{X =_{\nu} X\land Y}(\left\{Y \mapsto \bot \right\}) \\
      &= \finset{X \mapsto \bot, Y \mapsto \bot}
\end{align*}
\qed
\end{example}

\begin{definition}
\label{def:val_muclp}
A validity checking problem $(\phi,\mathcal{P})$ of a query $\phi$ for a \MuCLP{} $\mathcal{P}$ is that of deciding $\mathcal{P} \models \phi$, which we will also write $\models (\phi, \mathcal{P})$.
\end{definition}

\begin{remark}
\label{rem:CLP_MuCLP}
The validity of \MuCLP{} $(\phi_0,((X_1(\myseq{x}_1) =_{\alpha_1} \phi_1);\cdots;(X_m(\myseq{x}_m) =_{\alpha_m} \phi_m)))$ has an equivalent \CLP{} validity if all the following conditions are met: (1) $\alpha_i=\nu$ for all $i=1,\dots,m$, (2) $X_1,\dots,X_m$ occur only positively in $\phi_0$, and (3) universal (resp. existential) quantifiers occur only positively (resp. negatively) in $\phi_i$'s.  We henceforth call this fragment of \MuCLP{}, \emph{validity-reducible}.
Similarly, the \emph{invalidity} of \MuCLP{} has an equivalent \CLP{} validity if: (1) $\alpha_i=\mu$ for all $i=1,\dots,m$, (2) $X_1,\dots,X_m$ occur only positively in $\phi_0$, (3a) $\phi_i$'s only free variables are $\myseq{x}_i$, and (3b) universal (resp. existential) quantifiers occur only negatively (resp. positively) in $\phi_i$'s.  We call this \MuCLP{} fragment, \emph{invalidity-reducible}.
\qed
\end{remark}

%% file: apps_temp.tex
We now demonstrate the expressiveness of \MuCLP{} by showing that it can encode temporal property verification.  
In recent years, a wide variety of techniques and tools have emerged for verifying temporal properties of programs.  Here are some examples.  In the setting of infinite-state imperative programs, there have been works that prove CTL properties~\cite{Cook2011,Cook2013,Beyene2013}, LTL properties~\cite{Cook2011a,Dietsch2015}, and others such as CTL$^{*}$ properties~\cite{Cook2015a}.  For infinite-state higher-order programs, \cite{Murase2016} and \cite{Koskinen2014} respectively present automata-theoretic and type-based approaches to verification of $\omega$-regular properties (that subsume LTL).  As already mentioned, there are recent proposals of reductions from temporal program verification to validity checking in fixpoint logic~\cite{Kobayashi2018,Watanabe2019,Nanjo2018,Kobayashi2019}.  Our validity checking method for \MuCLP{} can be combined with their reductions to yield an automated temporal verification method for infinite-state imperative and functional programs that can solve classes of verification problems beyond the reach of the existing verification tools.

As an exemplary instance of such a reduction, we next formalize the
reduction from linear temporal property verification of infinite-state
systems to \MuCLP{}.
First, we review the notion of {\em labeled transition system} (LTS).
A LTS is a triple $\lts = (S,T,\labels)$ where $S \subseteq \domset^n$
is the the set of {\em states}, $\labels$ is the finite set of {\em
  labels}, $T \subseteq S \times\labels\times S$ is the {\em
  transition relation}.  (Note that $S$ may be infinite and therefore
we allow infinite-state systems.) For $\lts = (S,T,\labels)$, we often
write $S_\lts$ for $S$, $T_\lts$ for $T$, and $\labels_\lts$ for
$\labels$.  We write $s \lstep_\lts s'$ when $(s,\ell,s') \in T_\lts$.
We omit the subscript $\lts$ when it is clear from the context.

We now review the notion of a {\em B\"{u}chi automaton}.  A
(non-deterministic) B\"{u}chi automaton $\buchi$ is a tuple
$(Q,\labels,\delta,q_{\it init},F)$ where $Q$ is the finite set of
states (unrelated to the states of the LTS),
$\labels$ is the finite set of
labels, $\delta \subseteq Q\times \labels \times Q$ is the transition
relation, $q_{\it init} \in Q$ is the starting state, and
$F \subseteq Q$ is the set of final states.  For $q \in Q$ and
$\ell \in \labels$, we write $\delta(q,\ell)$ for the set
$\{ q'\in Q \mid (q,\ell,q') \in \delta \}$.  An infinite word
$\ell_0 \ell_1 \dots \in \labels^\omega$ is accepted by $\buchi$ if
and only if there exists an infinite sequence of states
$q_0,q_1,\dots$ such that $q_0 = q_{\it init}$,
$q_{i+1} \in \delta(q_i,\ell)$ for all $i \geq 0$, and some state in $F$
occurs infinitely often.

We consider the temporal property verification problem in which we are
given a LTS $\lts$, a predicate $\phi_{\it init}(\myseq{x})$ on states
of the LTS, and a B\"{u}chi automaton $\buchi$ such that the label set
of $\buchi$ is $\labels_\lts$.  Recall
that we allow a LTS to be
infinite-state.)  The goal of the verification is to decide if for any
(infinite) execution of $\lts$ from a state satisfying
$\phi_{\it init}(\myseq{x})$, the infinite sequence of labels of the
execution is accepted by $\buchi$.  That is, the goal is to verify
whether the given LTS satisfies the linear temporal property specified
by the given B\"{u}chi automaton.
The problem can be expressed in \MuCLP{} by defining the mutually-recursive
least-and-greatest fixpoint predicates $\ltlv_{q,\alpha}(\myseq{x})$ 
for each $q \in Q$ and $\alpha \in \{\mu,\nu\}$:
\[
\ltlv_{q,\alpha}(\myseq{x}) =_\alpha \bigwedge_{\ell \in \labels} \forall\myseq{y}.\tuple{\myseq{x}} \lstep \tuple{\myseq{y}} \Rightarrow \bigvee_{q' \in \delta(q,\ell)} \ltlv_{q',\alpha(q')}(\myseq{y}).
\]
Here, $\alpha(q) = \nu$ if $q \in F$ and $\alpha(q) = \mu$ otherwise.
Then, the LTS satisfies the temporal property if and only if
$\phi_{\it init}(\myseq{x}) \Rightarrow \ltlv_{q_{\it
    init},\nu}(\myseq{x})$ is valid.  The correctness of the
construction follows from the fact that $\ltlv_{q,\_}(\myseq{x})$
represents the set of states from which the labels along the execution
of the LTS is accepted by $\buchi$ when $\buchi$ is run from the state
$q$.  Note that the occurrence of a predicate in the body of the
recursive definition becomes the $\ltlv_{\_,\mu}$ variant when no
state in $F$ is visited in the corresponding execution step.  This
ensures that there must be a path in which a state from $F$ is visited
infinitely often.

%% file: pcsp.tex
We now describe a new verification intermediate representation, that generalizes \CHCS{}, and serves as an intermediary for automating \MuCLP{} validity queries.  Specifically, we formalize the class \PCSPWFFN{} of predicate constraint satisfaction problems.  We use $\varphi$ as a meta-variable ranging over $\theory$-formulas (cf.~Section~\ref{sec:muclp}) without quantifiers and predicate variables (but possibly with non-predicate function variables whose return sort is not $\propos$).  First, we define a \PCSP{} $\mathcal{C}$ (without function variables and well-founded predicate variables) to be a finite set of clauses of the form
\[
\varphi \lor \left( \bigvee_{i=1}^{\ell} X_i(\myseq{t}_i) \right) \lor \left( \bigvee_{i=\ell+1}^m \neg X_i(\myseq{t}_i) \right)
\]
where $0 \leq \ell \leq m$ and $\ffv{\varphi}=\emptyset$.

We write $\ftv{c}$ and $\ftv{\clauses}$ for the set of free term variables that occur in $c$ and $\clauses$, respectively.  We regard the variables in $\ftv{c}$ as implicitly universally quantified.  We write $\fpv{\clauses}$ (resp. $\ffv{\clauses}$) for the set of free predicate (resp. function) variables that occur in $\clauses$.  A \PCSP{} $\clauses$ is called \CHCS{} if $\ell \leq 1$ for all clauses $c \in \clauses$, and \COCHCS{} if $m \leq \ell+1$ for all $c \in \clauses$.  A \PCSP{} $\clauses$ is called linear \CHCS{} (or linear \COCHCS{}) if $\clauses$ is both \CHCS{} and \COCHCS{}.
A \emph{function/predicate substitution} $\sigma$ is a finite map from non-predicate function variables $F$ to closed functions of the form $\lambda x_1,\dots,x_{\arity{F}}.t$ and predicate variables $X$ to closed predicates of the form $\lambda x_1,\dots,x_{\arity{X}}.\varphi$.  We write $\sigma(\clauses)$ for the application of $\sigma$ to $\clauses$ and $\dom{\sigma}$ for the domain of $\sigma$.  We call $\sigma$ a \emph{syntactic solution} for $\clauses$ if $\ffv{\clauses} \subseteq \dom{\sigma}$ and $\models \bigwedge \sigma(\clauses)$.  Similarly, we call a function/predicate interpretation $\rho$ a \emph{semantic solution} for $\clauses$ if $\ffv{\clauses} \subseteq \dom{\rho}$ and $\rho \models \bigwedge \clauses$.

We next extend \PCSP{} to \PCSPWFFN{} by adding function variables and well-foundedness constraints.  A \PCSPWFFN{} $(\clauses,\WFRs)$ consists of
\begin{itemize}
    \item a finite set $\clauses$ of \PCSP{}-clauses over function/predicate variables without necessarily satisfying the \PCSP{}-restriction $\ffv{\varphi}=\emptyset$ of the $\varphi$-part of each clause and
    \item a set $\WFRs$ of well-founded predicate variables that are required to represent well-founded relations.
\end{itemize}
We write $\rho \models \WF{X}$ if the interpretation $\rho(X)$ of the predicate variable $X$ is \emph{well-founded}, that is, $\sort{X}=(\myseq{s},\myseq{s}) \to \propos$ for some sequence $\myseq{s}$ of sorts and there is no infinite sequence $\myseq{v}_1,\myseq{v}_2,\dots$ of sequences $\myseq{v}_i$ of values
of the sorts $\myseq{s}$ such that $(\myseq{v}_i,\myseq{v}_{i+1}) \in \rho(X)$ for all $i\geq 1$.
We call a function/predicate interpretation $\rho$ a \emph{semantic solution} for $(\clauses,\WFRs)$ if $\rho$ is a semantic solution of $\clauses$ and $\rho \models \mathit{WF}(X)$ for all $X \in \WFRs$.
The notion of syntactic solution can be similarly generalized to \PCSPWFFN{}.

\begin{definition}[Satisfiability of \PCSPWFFN{}]
\label{def:psat}
The predicate satisfiability problem of a \PCSPWFFN{} $(\clauses,\WFRs)$ is 
that of deciding whether it has a semantic solution.
\end{definition}

It is well known that the satisfiability of \CHCS{} and the validity of \CLP{} are inter-reducible.  In Section~\ref{sec:reduct}, we will show a sound and complete reduction from the validity of \MuCLP{} to the satisfiability of \PCSPWFFN{}.  The reduction is of practical importance because the latter problem is often easier to address: we may find a certificate of the satisfiability instead of exhaustively checking all possible cases.

%% file: reduct.tex
This section defines our reduction algorithm from the given \MuCLP{} validity problem $(\phi,\mathcal{P})$ (cf.~Definition~\ref{def:val_muclp}) to a \PCSPWFFN{} satisfiability problem $(\clauses,\WFRs)$ (cf.~Definition~\ref{def:psat}).
We assume without loss of generality that the (co-)inductive predicates $X \in \dom{\mathcal{P}}$ occur only positively in the query $\phi$: we can always transform the given query into this form by replacing each negative occurrence of $X$ in $\phi$ with $\neg X^\neg$ where the predicate $X^\neg$ is defined by the De Morgan dual $\neg \mathcal{P}$.

Our reduction consists of three steps: The first step, $\elimex$, Skolemizes positive occurrences of existential quantifiers and negative occurrences of universal quantifiers by introducing fresh function variables.  The second step, $\elimmu$, replaces inductive predicates defined by $\mu$-equations with co-inductive predicates defined by $\nu$-equations with guards (i.e., well-foundedness constraints) for co-recursion added to preserve the semantics.  The third step, $\elimnu$, further eliminates co-inductive predicates by replacing them with uninterpreted predicates represented as fresh predicate variables.  Formally, the reduction algorithm is:
\begin{align*}
\reduct{\phi,\mathcal{P}}
\defeq
\begin{array}{l}
\letmath\ (\phi_\mu,\mathcal{P}_\mu)
= \elimex(\phi,\mathcal{P}) \ \inmath \\
\letmath\ (\phi_\nu,\mathcal{P}_\nu,\WFRs)
= \elimmu(\phi_\mu,\mathcal{P}_\mu,\emptyset) \ \inmath (\elimnu(\phi_\nu,\mathcal{P}_\nu),\WFRs)
\end{array}
\end{align*}
Here, the \MuCLP{} $(\phi_\mu,\mathcal{P}_\mu)$ is obtained from $(\phi,\mathcal{P})$ by eliminating existential quantifiers with fresh function variables as stated above.  The definition of $\elimnu(\phi,\mathcal{P})$ is given as:
\begin{align*}
\elimnu (\phi,\epsilon) &\defeq \cnfOf{\uapprox{\phi}} \\
\elimnu (\phi,\mathcal{P};(X(\myseq{x})=_{\nu} \phi')) &\defeq
  \elimnu(\phi,\mathcal{P}) \cup \cnfOf{\uapprox{X}(\myseq{x}) \imply \uapprox{\phi'}}
\end{align*}
where $\uapprox{\phi}$ is the formula obtained from $\phi$ by replacing each occurrence of a predicate $X \in \dom{\mathcal{P}}$ with the predicate variable $\uapprox{X}$ that represents an under-approximation of $X$.  $\cnfOf{\phi}$ converts $\phi$ into its prenex and conjunctive normal form $\forall \myseq{x}.\bigwedge \clauses$ and returns the set $\clauses$ of clauses.  The most tricky part of the algorithm, namely, $\elimmu(\phi,\mathcal{P},\WFRs)$, is defined by:
\begin{align*}
  &\elimmu(\phi,(X_i(\myseq{x}_i)=_\nu \phi_i)_{i=1}^m,\WFRs)
  \defeq (\phi,(X_i(\myseq{x}_i)=_\nu \phi_i)_{i=1}^m,\WFRs) & \mbox{(base)} \\
  &\elimmu(\phi,\mathcal{P};(X(\myseq{x}) =_\mu \phi');(X_i(\myseq{x}_i)=_\nu \phi_i)_{i=1}^m, \WFRs)
  \defeq & \mbox{(recursive)} \\
  &\quad\elimmu(\sigma_0(\phi),\sigma_0(\mathcal{P});(X(\myseq{x}) =_\nu \sigma_X(\phi'));(X_i(b_i,\myseq{x},\myseq{x}_i)=_\nu \sigma_i(\phi_i))_{i=1}^m , \WFRs \cup \finset{\wfp{X}}) & \\
&\sigma_0 \defeq \myset{X_i \mapsto \lambda \myseq{y}.X_i(\bot,\myseq{v},\myseq{y})}{i=1,\dots,m} & \\
&\sigma_X \defeq \finset{X \mapsto \lambda \myseq{y}.X(\myseq{y}) \land \wfp{X}(\myseq{x},\myseq{y})} \cup \myset{X_i \mapsto \lambda \myseq{y}.X_i(\top,\myseq{x},\myseq{y})}{i=1,\dots,m} & \\
&\sigma_i \defeq \finset{X \mapsto \lambda \myseq{y}.X(\myseq{y}) \land (b_i \imply \wfp{X}(\myseq{x},\myseq{y}))} \cup \myset{X_j \mapsto \lambda \myseq{y}.X_j(b_i,\myseq{x},\myseq{y})}{j=1,\dots,m} &
\end{align*}
The third argument of $\elimmu$ accumulates generated fresh well-founded predicate variables.  The base case of $\elimmu(\phi,\mathcal{P},\WFRs)$ just returns the converted $\nu$-only \MuCLP{} $(\phi,\mathcal{P})$ that contains well-founded predicate variables in $\WFRs$.  In the recursive step, for the definition $X(\myseq{x})=_{\mu} \phi'$ of the right-most inductive (i.e., $\mu$) predicate $X$ in the input \MuCLP{}, we generate a fresh well-founded predicate variable $\wfp{X}$ and use it as the guard for each co-recursion in the converted co-inductive definition $X(\myseq{x}) =_\nu \sigma_X(\phi')$: we use the substitution $\sigma_X$ to replace each call $X(\myseq{t})$ in the body $\phi'$ of $X$ with $X(\myseq{t}) \land \wfp{X}(\myseq{x},\myseq{t})$ that checks that the formal arguments $\myseq{x}$ of $X$ and the actual arguments $\myseq{t}$ of the co-recursion are related by the well-founded relation represented by $\wfp{X}$.
At the same time, we extend the formal arguments of each co-inductive (i.e., $\nu$) predicate $X_i$ in the right-hand side of the equation for $X$ with arguments $\myseq{x}$ of the same sort as the formal arguments of $X$ and a Boolean argument $b_i$, where we assume that the formal arguments $\myseq{x}_i$ of $X_i$ are $\alpha$-renamed to avoid a name conflict between $\myseq{x}_i$ and $\myseq{x},b_i$.
The extended formal arguments $\myseq{x}$ of $X_i$ are used to receive the actual arguments previously passed to a call to the inductive predicate $X$ and are related by $\wfp{X}$, in the converted definition of $X_i$, with the actual arguments passed to each indirect recursive call to $X$ in $X_i$.\footnote{This transformation is similar in spirit to binary reachability analysis~\cite{Cook2006,Podelski2004a,Kuwahara2014} for termination verification.}  
Dummy values are passed as $\myseq{x}$ when no such previous call to $X$ exists and the extended Boolean formal argument $b_i$ of $X_i$ indicates whether there indeed is such a call to $X$ and its actual arguments are passed as $\myseq{x}$ to $X_i$ ($b_i=\top$) or the dummy values are passed as $\myseq{x}$ to $X_i$ ($b_i=\bot$).
In fact, we use the substitution $\sigma_0$ to replace each call $X_i(\myseq{t})$ in the query $\phi$ and the definition of the predicates $\mathcal{P}$ in the left-hand side of the equation for $X$ with $X_i(\bot,\myseq{v},\myseq{t})$ for some sequence $\myseq{v}$ of dummy values of the same sorts as the formal arguments $\myseq{x}$ of $X$.  For the definition $X(\myseq{x})=_{\mu} \phi'$, we use the substitution $\sigma_X$ to replace each call $X_i(\myseq{t})$ in $X$ with $X_i(\top,\myseq{x},\myseq{t})$.
For the definition $X_j(\myseq{x}_j)=_\nu \phi_j$ of each co-inductive predicate $X_j$ in the right-hand side of the equation for $X$, we use $\sigma_j$ to replace each call $X_i(\myseq{t})$ in $X_j$ with $X_i(b_j,\myseq{x},\myseq{t})$ and each call $X(\myseq{t})$ with $X(\myseq{t}) \land (b_j \imply \wfp{X}(\myseq{x},\myseq{t}))$ that checks that if $\myseq{x}$ are not dummy (i.e., $b_j = \top$), the actual arguments of a previous call to $X$ passed around to $X_j$ as its extended formal arguments $\myseq{x}$ are related by $\wfp{X}$ with the actual arguments $\myseq{t}$ of the indirect recursive call to $X$.  In the resulting \MuCLP{}, the generated well-founded predicate variables occur only positively.

\begin{example}
Let us consider the \MuCLP{} $(\phi,\mathcal{P})$ where
$\phi \defeq \forall x. X(x) \land Y(x)$ and
\begin{align*}
\mathcal{P} &\defeq (X(x) =_{\mu} Y(x-1)); (Y(y) =_{\mu} y \leq 0 \lor X(y-1))
\end{align*}
We obtain $ \elimex{(\phi,\mathcal{P})=(\phi,\mathcal{P})}$ and
\begin{align*}
\elimmu{(\phi,\mathcal{P},\emptyset)}
&=\elimmu{(\phi,(X(x) =_{\mu} Y(x-1)); (Y(y) =_{\nu} y \leq 0 \lor X(y-1)),\emptyset)} \\
&=\left(
\begin{array}{l}
\forall x. X(x) \land Y(\bot,0,x), \\
(X(x) =_{\nu} Y(\top,x,x-1)); \\
(Y(b,x,y) =_{\nu} y \leq 0 \lor X(y-1) \land (b \imply \wfp{X}(x,y-1))), \\
\finset{\wfp{X}}
\end{array}\right)
\end{align*}
Here, in the first step of the transformation, the inductive definition of $Y$ is simply replaced by the co-inductive definition because the body of $Y$ has no recursive call to $Y$.  The indirect recursive call to $X$ in $Y$ is properly handled in the second step by adding the formal arguments $b$ and $x$ to $Y$.  Note also that in the call $Y(\bot,0,x)$ in the query, $0$ is used as a dummy value for the extended formal argument $x$ of $Y$.

We thus get \PCSP{} $\reduct{\phi,\mathcal{P}}=(\clauses,\finset{\wfp{X}})$ where
\[\begin{array}{ll}
\clauses &\defeq \finset{\begin{array}{l}
\uapprox{X}(x),\ 
\uapprox{Y}(\bot,0,x),\ 
\neg \uapprox{X}(x) \lor \uapprox{Y}(\top,x,x-1), \\
\neg \uapprox{Y}(b,x,y) \lor y \leq 0 \lor \uapprox{X}(y-1),\ 
\neg \uapprox{Y}(b,x,y) \lor y \leq 0 \lor \neg b  \lor \wfp{X}(x,y-1))
\end{array}} \\
\end{array}\]
\qed
\end{example}
\vspace{-10pt}
\begin{remark}
In the implementation of $\reduct{\phi,\mathcal{P}}$ in our \MuCLP{} validity checker \muval{}, unnecessary arguments addition is suppressed.
For example, from the \MuCLP{} $(\phi,\mathcal{P})$ where
\begin{align*}
\phi &\defeq \forall x. X(x) \\
\mathcal{P} &\defeq (X(x) =_{\mu} Y(x-1)); (Y(y) =_{\mu} Z(y-1)); (Z(z) =_{\mu} z \leq 0 \lor X(z-1))
\end{align*}
$\reduct{\phi,\mathcal{P}}$ generates the \PCSP{} $(\clauses,\finset{\wfp{X}})$ where
\begin{align*}
\clauses &\defeq \finset{\begin{array}{l}
\uapprox{X}(x),\ 
\uapprox{X}(x) \imply \uapprox{Y}(\top,x,x-1),\ 
\uapprox{Y}(b_1,x,y) \imply \uapprox{Z}(b_1,x,\top,y,y-1), \\
\uapprox{Z}(b_1,x,b_2,y,z) \imply z \leq 0 \lor \uapprox{X}(z-1),\ 
\uapprox{Z}(b_1,x,b_2,y,z) \land b_1 \imply z \leq 0 \lor \wfp{X}(x,z-1))
\end{array}}
\end{align*}
By contrast, \muval{} gets a simpler but equi-satisfiable \PCSP{} $(\clauses',\finset{\wfp{X}})$ where
\[\begin{array}{ll}
\clauses' &\defeq \finset{\begin{array}{l}
\uapprox{X}(x),\ 
\uapprox{X}(x) \imply \uapprox{Y}(x,x-1),\ 
\uapprox{Y}(x,y) \imply \uapprox{Z}(x,y-1), \\
\uapprox{Z}(x,z) \imply z \leq 0 \lor \uapprox{X}(z-1),\ 
\uapprox{Z}(x,z) \imply z \leq 0 \lor \wfp{X}(x,z-1))
\end{array}}
\end{array}\]
Note that the query $\phi$ calls $X$, $X$ calls $Y$, $Y$ calls $Z$, and $Z$ recursively calls $X$.  Thus, if we start from the query, $Z$ is always passed the actual argument of the previous call to $X$ and, therefore, $b_1$ is always $\top$ and so unneeded.  Likewise, $b_2$ is also always $\top$ and unneeded.
\qed
\end{remark}

We now show the following soundness and the completeness of the reduction algorithm.
\begin{theorem}
\label{thm:red_sound_comp}
$\reduct{\phi,\mathcal{P}}$ has a semantic solution if and only if $\models (\phi,\mathcal{P})$.
\end{theorem}

This follows from the lemmas for each steps of $\reduct{\phi,\mathcal{P}}$: The following lemma for the first-step $\elimex(\phi,\mathcal{P})$ follows immediately from the well-known soundness and completeness of Skolemization for first-order logic.
\begin{lemma}
$\models (\phi,\mathcal{P})$ if and only if there is an interpretation $\rho$ for the function variables introduced by the Skolemization such that $\rho \models \elimex(\phi,\mathcal{P})$.
\end{lemma}

The following lemma for the third-step $\elimnu(\phi,\mathcal{P})$ follows from the maximality of the greatest fixpoints (i.e., co-induction principle) (see e.g., Corollary 1 in \cite{Unno2017b} for a formal related discussion of least fixpoints that occur in negative positions).
\begin{lemma}
Let $\rho$ be any interpretation of $\ffv{\phi,\mathcal{P}}$.
$\elimnu(\phi,\mathcal{P})$ has a semantic solution that extends $\rho$ if and only if $\rho \models (\phi,\mathcal{P})$.
\end{lemma}

We finally show the soundness and completeness of the second-step $\elimmu(\phi,\mathcal{P})$.
\begin{lemma}
Suppose that $\elimmu(\phi,\mathcal{P})=(\phi',\mathcal{P}',\mathcal{R})$.  We then have $\models (\phi,\mathcal{P})$ if and only if there is an interpretation $\rho$ of $\mathcal{R}$ such that $\rho \models \WF{X}$ for all $X \in \mathcal{R}$ and $\rho \models (\phi',\mathcal{P}')$.
\end{lemma}

This can be shown as a corollary of the following lemma.
\begin{lemma}
\label{lem:elimmu}
$\rho \models (\phi,\mathcal{P};(X(\myseq{x}) =_\mu \phi');(X_i(\myseq{x}_i)=_\nu \phi_i)_{i=1}^m)$ if and only if there is an interpretation $\rho'$ of $\wfp{X}$ such that $\rho' \models \WF{\wfp{X}}$ and $\rho \uplus \rho' \models (\sigma_0(\phi),\sigma_0(\mathcal{P});(X(\myseq{x}) =_\nu \sigma_X(\phi'));(X_i(b_i,\myseq{x},\myseq{x}_i)=_\nu \sigma_i(\phi_i))_{i=1}^m)$ where
\begin{align*}
\sigma_0 &\defeq \myset{X_i \mapsto \lambda \myseq{y}.X_i(\bot,\myseq{v},\myseq{y})}{i=1,\dots,m} & \\
\sigma_X &\defeq \finset{X \mapsto \lambda \myseq{y}.X(\myseq{y}) \land \wfp{X}(\myseq{x},\myseq{y})} \cup \myset{X_i \mapsto \lambda \myseq{y}.X_i(\top,\myseq{x},\myseq{y})}{i=1,\dots,m} & \\
\sigma_i &\defeq \finset{X \mapsto \lambda \myseq{y}.X(\myseq{y}) \land (b_i \imply \wfp{X}(\myseq{x},\myseq{y}))} \cup \myset{X_j \mapsto \lambda \myseq{y}.X_j(b_i,\myseq{x},\myseq{y})}{j=1,\dots,m} &
\end{align*}
\end{lemma}

\begin{remark}
\label{rem:CHC_coCHC_MuCLP}
Let $(\phi,\mathcal{P})$ be a \MuCLP{}.  If $(\phi,\mathcal{P})$ is validity-reducible (recall Remark~\ref{rem:CLP_MuCLP}), $\reduct{\phi,\mathcal{P}}$ always generates \COCHCS{}.  Similarly, if $(\phi,\mathcal{P})$ is invalidity-reducible, $\reduct{\phi,\neg(\mathcal{P})}$ always generates \COCHCS{}.  Note also that the satisfiability of the \COCHCS{} $\clauses$ can be further reduced to that of the \CHCS{} $\clauses^\neg$, obtained from $\clauses$ by replacing, each literal of the form $X(\myseq{t})$ with $\neg X^{\neg}(\myseq{t})$ and $\neg X(\myseq{t})$ with $X^{\neg}(\myseq{t})$ for $X \in \fpv{\clauses}$, where $X^\neg$ is a fresh predicate variable that represents the negation of $X$.
Thus we can use off-the-shelf CHC solvers to discharge the validity-reducible and invalidity-reducible fragments of \MuCLP{}.  Our constraint solving method described in the next section can handle the full classes of \PCSPWFFN{} and \MuCLP{} (via the reduction).
\qed
\end{remark}

%% file: csolve.tex
This section describes our CEGIS-based method for finding a (syntactic) solution---in other words, (co-)inductive invariants, ranking functions, and witnesses for existential quantifiers---of the given \PCSPWFFN{} $(\clauses,\WFRs)$.  Our method iteratively accumulates \emph{example instances of $\clauses$}, which are defined to be \PCSPWFFN{}-clauses without term variables obtained from $\clauses$ by instantiating $\ftv{\clauses}$, from which a sequence of candidate solutions for $\clauses$ is generated by using a synthesizer $\mathcal{S}$ (whose details are deferred to Section~\ref{sec:synth_tb}), until a genuine solution or a counterexample (i.e., unsatisfiable example instances) is found.  We write $\examples^{(i)}$ for the set of example instances accumulated before the iteration $i$.  Starting from $\examples^{(1)}=\emptyset$, for each iteration $i\geq1$, our method performs the following:
\begin{enumerate}
\item {\bf Synthesis Phase}: We check whether the set of instances $(\examples^{(i)},\WFRs)$ is unsatisfiable.  If so, we return $\examples^{(i)}$ as a counter example to the input \PCSPWFFN{} $(\clauses,\WFRs)$.  Otherwise, we let the synthesizer $\mathcal{S}$ find a syntactic solution $\sigma^{(i)}$ (with $\dom{\sigma^{(i)}}=\ffv{\clauses}$) of the instances $(\examples^{(i)},\WFRs)$, which will be used as a candidate solution for $(\clauses,\WFRs)$.
\item {\bf Validation Phase}: We check whether $\sigma^{(i)}$ is a genuine solution to $(\clauses,\WFRs)$ by using an off-the-shelf SMT solver.  If so, we return $\sigma^{(i)}$ as the solution.  Otherwise, for each clause $c \in \clauses$ unsatisfied by $\sigma^{(i)}$, we obtain a counterexample, that is, a term substitution $\theta_c$ such that $\dom{\theta_c}=\ftv{c}$ and $\not\models \theta_c(\sigma^{(i)}(c))$.  We then update the example set by adding a new example instance for each unsatisfied clause (i.e., $\examples^{(i+1)}=\examples^{(i)} \cup \myset{\theta_c(c)}{c \in \clauses \land \not\models \sigma^{(i)}(c)}$), and proceed to the next iteration with $\examples^{(i+1)}$.
\end{enumerate}
\begin{remark}
In our \PCSPWFFN{} satisfiability checker \pcsat{} (Section~\ref{sec:eval}), we implemented a third phase that we call the \emph{resolution phase} for accelerating the convergence of the CEGIS loop.  There, we first apply unit propagation repeatedly to the example instances $\examples$ to obtain the set $\examples^+$ of positive examples of the form $X(\myseq{v})$ and the set $\examples^-$ of negative examples of the form $\neg X(\myseq{v})$.  We then repeatedly apply resolution principle to the clauses in the original \PCSPWFFN{} $\clauses$ and the clauses in $\examples^+ \cup \examples^-$ to obtain new positive/negative examples (without containing term variables), which are then added to $\examples$.
\end{remark}

In general, the above CEGIS procedure may diverge, which is inevitable due to the undecidability of \PCSPWFFN{}.  But it satisfies the so called \emph{progress property}: any counterexample and candidate solution found in an iteration are never generated again in succeeding iterations.  Furthermore, if we carefully design a synthesizer $\mathcal{S}$ as discussed in Section~\ref{sec:synth_tb} by incorporating our idea of stratified CEGIS, we can show the \emph{relative completeness} in the sense of \cite{Jhala2006,Terauchi2015}: if the given \PCSPWFFN{} $(\clauses,\WFRs)$ has a \emph{syntactic} solution expressible in the
stratified families of templates, a solution of the \PCSPWFFN{} is eventually found by the procedure.

The rest of this section discusses the details of the synthesis phase.  Section~\ref{sec:unsat_ex} discusses how to check the unsatisfiability of example instances $(\examples,\WFRs)$.  Section~\ref{sec:synth_tb} discusses the synthesis based on stratified template families and unsat-core-based template refinement.  For simplicity, we focus on the theory of quantifier-free linear integer arithmetic (QFLIA) in the description of the synthesis phase.  Designing stratified template families for richer theories such as arrays, algebraic data types, and heaps is actually non-trivial and will be discussed as a future work in Section~\ref{sec:conc}.

\subsection{Unsatisfiability Checking of Example Instances}
\label{sec:unsat_ex}
\input{unsat_ex}

\subsection{Function/Predicate Synthesis with Stratified Families of Templates}
\label{sec:synth_tb}
\input{synth_tb}

%% file: unsat_ex.tex
If $\WFRs=\emptyset$, the unsatisfiability of the given example instances $(\examples,\WFRs)$ can be decided by an off-the-shelf SAT solver (if $(\ffv{\examples} \setminus \fpv{\examples})=\emptyset$) or SMT solver (otherwise) because $\examples$ is a finite set of clauses not containing term variables.  Otherwise, we use the following (CDCL-like) iterative algorithm staring from $\examples_0=\examples$: For each iteration $i \geq 0$, we first check whether $(\examples_i,\emptyset)$ is unsatisfiable.  If so, then we conclude that $(\examples,\WFRs)$ is unsatisfiable.  Otherwise, we obtain a satisfying assignment $\sigma$ for $\examples_i$.  Then, for each $X \in \WFRs$, we consider the graph comprising the edges $\myset{(\myseq{v}_1,\myseq{v}_2)}{\;\models \sigma(X(\myseq{v}_1,\myseq{v}_2))}$ and enumerate its simple cycles (e.g., by using the algorithm of \cite{Johnson1975}).  Note that such cycles would be counterexamples to the well-foundedness constraint $X$.
If no such cycles
exist, we conclude that $(\examples,\WFRs)$ is satisfiable.  Otherwise, we let $\examples_{i+1}$ be $\examples_i$ but with the following new learnt clauses added:
\begin{itemize}
    \item $\neg X(\myseq{v}_1,\myseq{v}_2) \lor \dots \lor \neg X(\myseq{v}_{m-1},\myseq{v}_m)$ for each simple cycle $\myseq{v}_1,\dots,\myseq{v}_m=\myseq{v}_1$ of each $X \in \WFRs$.
\end{itemize}
We then proceed to the next iteration with $\examples_{i+1}$.

It is worth mentioning here that if $\WFRs=\emptyset$ and the original \PCSP{} $(\clauses,\WFRs)$ is unsatisfiable, there always exists an unsatisfiable finite set $\examples$ of example instances of $\clauses$.
However, there is, in general, no such finite witness of the unsatifiability if $\WFRs \neq \emptyset$.  This fact also supports an advantage of our primal-dual approach to verification based on \MuCLP{}.

%% file: synth_tb.tex
We do a template-based search for a solution of the given example instances to be returned as a candidate solution of the input \PCSPWFFN{} $(\clauses,\WFRs)$.  Templates can effectively restrict the solution space to explore and be made to satisfy the well-foundedness constraints at the same time.
There however is a trade-off between expressiveness and generalizability.  With less expressive templates like intervals, we may miss actual solutions.  By contrast, with very expressive templates like polyhedra, there could be many solutions, and a solution thus returned is liable to overfitting and therefore is of low generalizability.  That is, the solution is likely too specific to be a solution of $(\clauses,\WFRs)$.  \cite{Padhi2019} discusses a similar overfitting problem in the context of grammar-based synthesis.

Our remedy to the problem is to use stratified families of predicate templates that have been used in prior work to guarantee the convergence of counterexample-guided refinement iterations~\cite{Jhala2006,Terauchi2015}.  Initially, we assign each predicate variable a less expressive template and gradually refine it in a counterexample-guided manner: we try to find a solution expressible in the current templates and if no solution is found, we generate and analyze an unsat core of the constraint over the unknown parameters of the templates to identify the \emph{parameters of the families of templates} that are necessary to be updated.

\subsubsection{Stratified Families of Templates}

We have designed three stratified families of templates respectively for (1) ordinary predicates, (2) (non-predicate) functions, and (3) well-founded predicates.
\begin{enumerate}
\item For ordinary predicates $X \in (\fpv{\clauses} \setminus \WFRs)$, the stratified family of templates $\template_X^\bullet(\mathit{nd},\mathit{nc},\mathit{ac},\mathit{ad})$ and its accompanying constraint $\phi_X^\bullet(\mathit{nd},\mathit{nc},\mathit{ac},\mathit{ad})$ are defined as:
\begin{align*}
\begin{array}{rcl}
\template_X^\bullet(\mathit{nd},\mathit{nc},\mathit{ac},\mathit{ad})
&\defeq&
\lambda (x_1,\dots,x_{\arity{X}}). \bigvee_{i=1}^{\mathit{nd}} \bigwedge_{j=1}^{\mathit{nc}} c_{i,j,0}+\sum_{k=1}^{\arity{X}} c_{i,j,k} \cdot x_k \geq 0 \\
\phi_X^\bullet(\mathit{nd},\mathit{nc},\mathit{ac},\mathit{ad}) &\defeq&
\bigwedge_{i=1}^{\mathit{nd}} \bigwedge_{j=1}^{\mathit{nc}}
\sum_{k=1}^{\arity{X}} \abs{c_{i,j,k}} \leq \mathit{ac} 
\land
\abs{c_{i,j,0}} \leq \mathit{ad}
\end{array}
\end{align*}
Here, $c_{i,j,k}$'s are fresh unknown parameters to be inferred.  Note that the parameter $\mathit{nd}$ (resp. $\mathit{nc}$) is the number of disjuncts (resp. conjuncts) and the parameter $\mathit{ac}$ is the upper bound of the sum of the absolute value of coefficients $c_{i,j,k}\ (k > 0)$ of the variables.  The parameter $\mathit{ad}$ is the upper bound of the absolute value of constant term $c_{i,j,0}$.

\item For (non-predicate) functions $F \in (\ffv{\clauses} \setminus \fpv{\clauses})$, we define the stratified family of templates $\template_F^\lambda(\mathit{nd},\mathit{nc},\mathit{dc},\mathit{dd},\mathit{ec},\mathit{ed})$ and its accompanying constraint $\phi_F^\lambda(\mathit{nd},\mathit{nc},\mathit{dc},\mathit{dd},\mathit{ec},\mathit{ed})$ as:
\begin{align*}
\begin{array}{rcl}
\template_F^\lambda(\mathit{nd},\mathit{nc},\mathit{dc},\mathit{dd},\mathit{ec},\mathit{ed}) &\defeq&
\lambda (\myseq{x}).t_1(\myseq{x})
\\
\phi_F^\lambda(\mathit{nd},\mathit{nc},\mathit{ec},\mathit{ed},\mathit{dc},\mathit{dd}) &\defeq&
\begin{array}{l}
\bigwedge_{i=1}^{\mathit{nd}}
\sum_{j=1}^{\arity{X}-1} \abs{c_{i,j}} \leq \mathit{ec}
\land
\abs{c_{i,0}} \leq \mathit{ed}\ \land  \\
\bigwedge_{i=1}^{\mathit{nd}-1}
\bigwedge_{j=1}^{\mathit{nc}}
\sum_{k=1}^{\arity{X}-1} \abs{c_{i,j,k}'} \leq \mathit{dc}
\land
\abs{c_{i,j,0}'} \leq \mathit{dd}
\end{array}
\end{array}
\end{align*}
where
\begin{align*}
\begin{array}{rclrcl}
t_{\mathit{nd}}(\myseq{x}) &\defeq& e_{\mathit{nd}}(\myseq{x}), &
t_i(\myseq{x}) &\defeq& \ITE(D_i(\myseq{x}),e_i(\myseq{x}),t_{i+1}(\myseq{x}))\quad(\mbox{for } 1 \leq i < \mathit{nd}), \\
e_{i}(\myseq{x}) &\defeq& c_{i,0} + \sum_{j=1}^{\arity{X}-1} c_{i,j} \cdot x_j, &
D_{i}(\myseq{x}) &\defeq& \bigwedge_{j=1}^{\mathit{nc}} c_{i,j,0}' + \sum_{k=1}^{\arity{X}-1} c_{i,j,k}' \cdot x_k \geq 0.
\end{array}
\end{align*}
Here, $c_{i,j}$'s and $c_{i,j,k}'$'s are fresh unknown parameters to be inferred.  $\template_F^\lambda$ characterizes a piecewise-defined affine function with discriminators $D_1,\dots,D_{\mathit{nd}-1}$ and branch expressions $e_1,\dots,e_{\mathit{nd}}$. The parameter $\mathit{nc}$ is the number of conjuncts in each discriminator.
The parameters $\mathit{dc},\mathit{dd},\mathit{ec},\mathit{ed}$ are the upper bounds similar to 
$\mathit{ac},\mathit{ad}$ for $\template_X^\bullet$.
Note that for any substitution $\theta$ for the unknown parameters in $\template_F^\lambda$, $\theta(\template_F^\lambda)$ represents a total function.

\item For well-founded predicates $X \in \WFRs$, the stratified family of templates $\template_X^\Downarrow(\mathit{nl},\mathit{np},\mathit{nc},\mathit{rc},\mathit{rd},\mathit{dc},\mathit{dd})$ and its accompanying constraint $\phi_X^\Downarrow(\mathit{nl},\mathit{np},\mathit{nc},\mathit{rc},\mathit{rd},\mathit{dc},\mathit{dd})$ are defined as:
\begin{align*}
\begin{array}{rcl}
\template_X^\Downarrow(\mathit{nl},\mathit{np},\mathit{nc},\mathit{rc},\mathit{rd},\mathit{dc},\mathit{dd})
&\defeq&
\lambda (\myseq{x},\myseq{y}).
\begin{array}{l}
\left(\bigwedge_{i=1}^{\mathit{nl}}\bigwedge_{j=1}^{\mathit{np}} r_{i,j}(\myseq{x}) \geq 0\right)\ \land \\
\left(\bigwedge_{i=1}^{\mathit{nl}}\bigvee_{j=1}^{\mathit{np}} D_{i,j}(\myseq{x})\right) \land
\left(\bigwedge_{i=1}^{\mathit{nl}}\bigvee_{j=1}^{\mathit{np}} D_{i,j}(\myseq{y})\right)\ \land \\
\left(\bigvee_{i=1}^{\mathit{nl}}
\mathit{GT}_i(\myseq{x},\myseq{y}) \land \bigwedge_{\ell=1}^{i-1} \mathit{GEQ}_\ell(\myseq{x},\myseq{y})\right)
\end{array}
 \\
\phi_X^\Downarrow(\mathit{nl},\mathit{np},\mathit{nc},\mathit{rc},\mathit{rd},\mathit{dc},\mathit{dd})
&\defeq&
\begin{array}{l}
\bigwedge_{i=1}^{\mathit{nl}}
\bigwedge_{j=1}^{\mathit{np}}
\sum_{k=1}^{\arity{X}} \abs{c_{i,j,k}} \leq \mathit{rc} \land
\abs{c_{i,j,0}} \leq \mathit{rd}\ \land \\
\bigwedge_{i=1}^{\mathit{nl}}
\bigwedge_{j=1}^{\mathit{np}}
\bigwedge_{k=1}^{\mathit{nc}}
\sum_{l=1}^{\arity{X}} \abs{c_{i,j,k,l}'} \leq \mathit{dc} \land
\abs{c_{i,j,k,0}'} \leq \mathit{dd}
\end{array}
\end{array}
\end{align*}
where
\begin{align*}
\begin{array}{rcl}
\mathit{GT}_i(\myseq{x},\myseq{y}) &\defeq&
\bigvee_{j=1}^{\mathit{np}}
D_{i,j}(\myseq{x})\geq0 \land
\bigwedge_{k=1}^{\mathit{np}}
\left(D_{i,k}(\myseq{y})\geq0 \imply r_{i,j}(\myseq{x}) > r_{i,k}(\myseq{y})\right) \\
\mathit{GEQ}_i(\myseq{x},\myseq{y}) &\defeq&
\bigvee_{j=1}^{\mathit{np}}
D_{i,j}(\myseq{x})\geq0 \land
\bigwedge_{k=1}^{\mathit{np}}
\left(D_{i,k}(\myseq{y})\geq0 \imply r_{i,j}(\myseq{x}) \geq r_{i,k}(\myseq{y})\right) \\
D_{i,j}(\myseq{x}) &\defeq&
\bigwedge_{k=1}^{\mathit{nc}}
c_{i,j,k,0}' + \sum_{l=1}^{\arity{X}/2} c_{i,j,k,l}' \cdot x_l \geq 0 \\
r_{i,j}(\myseq{x}) &\defeq&
c_{i,j,0} + \sum_{k=1}^{\arity{X}/2} c_{i,j,k} \cdot x_k
\end{array}
\end{align*}
Here, $c_{i,j,k}$'s and $c_{i,j,k,l}'$'s are fresh unknown parameters to be inferred.  $\template_X^\Downarrow$ represents the well-founded relation induced by the $\mathit{nl}$-lexicographic $\mathit{np}$-piecewise-defined affine ranking function where $r_{i,j}$ are the affine ranking function template for the $j$-th region specified by the discriminator $D_{i,j}$ of the $i$-th lexicographic component.  The parameters $\mathit{rc},\mathit{rd},\mathit{dc},\mathit{dd}$ are the upper bounds similar to 
$\mathit{ac},\mathit{ad}$ for $\template_X^\bullet$.
The first conjunct of $\template_X^\Downarrow$ asserts that the return value of all the affine ranking functions is non-negative and the second conjunct asserts that the discriminators cover all the reachable states.  Note that discriminators can overlap and for such an overlapping region, the maximum return value of the ranking functions is used.  In the third conjunct, $\mathit{GT}_i(\myseq{x},\myseq{y})$ (resp. $\mathit{GEQ}_i(\myseq{x},\myseq{y})$) means that the return value of the piecewise-defined affine ranking function of $i$-th lexicographic component strictly (resp. non-strictly) decreases from $\myseq{x}$ to $\myseq{y}$.
It follows that for any substitution $\theta$ for the unknown parameters in $\template_X^\Downarrow$, $\theta(\template_X^\Downarrow)$ represents a well-founded relation.
\end{enumerate}

\subsubsection{Template-based Synthesis}

For each function variable $F$, let $\mathit{ty}(F)=\bullet$ if $F$ is ordinary predicate, $\mathit{ty}(F)=\lambda$ if it is (non-predicate) function, and $\mathit{ty}(F) = \mathord{\Downarrow}$ if it is well-founded predicate.  Let $\myseq{p} \in \mathbb{Z}^n$ where $n$ is the number of parameters summed across all templates, and let $\template_F^\alpha(\myseq{p})$ and $\phi_F^\alpha(\myseq{p})$ (for $\alpha \in \finset{\bullet, \lambda, \Downarrow}$) project the corresponding parameters.  Each $\myseq{p} \in \mathbb{Z}^n$ induces a solution space $\denote{}{\myseq{p}} \defeq \{ \template(\myseq{p})[\theta] \mid \theta \models \mathit{Con}(\myseq{p})\}$ where $\template(\myseq{p})[\theta] \defeq \{ F \mapsto \theta(\template_F^{\mathit{ty}(F)}(\myseq{p})) \mid F \in \ffv{\mathcal{C}}\}$ and $\mathit{Con}(\myseq{p}) \defeq \bigwedge_{F \in \ffv{\mathcal{C}}} \phi_F^{\mathit{ty}(F)}(\myseq{p})$.

Let $\myseq{p_1} \leq \myseq{p_2}$ be the point-wise ordering.  Note that $\denote{}{\myseq{p}}$ is a finite set for any $\myseq{p} \in \mathbb{Z}^n$, and $\myseq{p_1} \leq \myseq{p_2}$ implies $\denote{}{\myseq{p}_1} \subseteq \denote{}{\myseq{p}_2}$.
We start the CEGIS process with some small initial parameters $\myseq{p}^{(0)}$ (i.e., the parameters will be maintained as a {\em global state} of the CEGIS process).  At each iteration, we try to find a candidate solution to the given examples $(\examples,\WFRs)$ in $\denote{}{\myseq{p}^{(i)}}$ where $\myseq{p}^{(i)}$ are the current parameters.  This is done by using an off-the-shelf SMT solver for QFLIA to find $\theta$ satisfying $\bigwedge \template(\myseq{p}^{(i)})[\theta](\examples) \wedge \theta(\mathit{Con}(\myseq{p}^{(i)}))$.  If such $\theta$ is found, we return $\template(\myseq{p}^{(i)})[\theta]$ as the candidate solution for the outer CEGIS process.  Otherwise, we update the parameters to some $\myseq{p}^{(i+1)} > \myseq{p}^{(i)}$ such that $\denote{}{\myseq{p}^{(i+1)}}$ contains a solution for $(\examples,\WFRs)$.  Here, we do the update in a {\em fair manner}, that is, in any infinite series of updates $\myseq{p}^{(0)},\myseq{p}^{(1)},\dots$, every parameter is updated infinitely often (the details are deferred to below).  Because of the progress property of our CEGIS procedure (cf.~Section~\ref{sec:csolve}) and the fact that every $\denote{}{\myseq{p}}$ is finite, this ensures that every parameter is updated infinitely often in an infinite series of CEGIS iterations.  We thus obtain the following property.
\begin{theorem}
\label{thm:relcomp}
Our CEGIS-procedure based on stratified families of templates is relatively complete: if there is $\myseq{p}$ and $\sigma \in \denote{}{\myseq{p}}$ such that $\sigma$ is a syntactic solution of the given \PCSPWFFN{} $(\clauses,\WFRs)$, a syntactic solution $\sigma'$ of $(\clauses,\WFRs)$, which could be different from $\sigma$, is eventually found by the procedure.
\end{theorem}

\subsubsection{Updating Template Parameters via Unsat Cores}
In above, when the constraint $\bigwedge \template(\myseq{p}^{(i)})[\theta](\examples) \wedge \theta(\mathit{Con}(\myseq{p}^{(i)}))$ has no solution (and thus $\denote{}{\myseq{p}^{(i)}}$ has no solution to the examples), we analyze the {\em unsat core} of the constraint to obtain the parameters that have caused the failure.
Note here that there could be a dependency between function/predicate variables and in such a case our unsat core analysis enumerates all the involved function/predicate variables and we obtain the parameters of the templates for all of them.
We then increment these parameters in a fair manner, by limiting the maximum differences between different parameters to some finite threshold, and repeatedly solve the resulting constraint until a solution is found.

%% file: eval.tex
To evaluate the presented verification framework, we have implemented:
\begin{itemize}
    \item \pcsat{}, a satisfiability checking tool for \PCSPWFFN{} based on stratified CEGIS.
    \item \muval{}, a validity checking tool for \MuCLP{} based on the reduction algorithm presented in Section~\ref{sec:reduct} and the satisfiability checker \pcsat{}.
\end{itemize}
\pcsat{} supports the theory of Booleans and the quantifier-free theory of linear inequalities over integers/rationals.  The tools are implemented in OCaml, using Z3~\cite{Moura2008} and MiniSat~\cite{Een2004} as the backend SMT and SAT solvers, respectively.

We compare \pcsat{} with the state-of-the-art SyGuS (syntax-guided synthesis) solver \loopinvgen{}~\cite{Padhi2019} which is the winner of the Inv Track of SyGuS-Comp 2018.  We also compare with the state-of-the-art \CHCS{} solvers \hoice{}~\cite{Champion2018} and \spacer{}~\cite{Gurfinkel2015}.  We run the tools on the following benchmark sets:
\begin{itemize}
    \item[(a)] SyGuS-Comp 2018 (Invariant Synthesis Track).
    \item[(b)] CHC-COMP 2019 (LIA-nonlin Track) for \CHCS{} over the theory of QFLIA.
\end{itemize}
We remark that the SyGuS benchmarks only contain linear \CHCS{} with each constraint set containing only a single predicate variable.  To compare, we have selected non-linear instances from CHC-COMP.

We have also tested \muval{} on the benchmark sets below encoded as \MuCLP{} and compared the results with \mutochc{}~\cite{Kobayashi2019}, which is a recently proposed tool for solving fixpoint logic constraints:
\begin{itemize}
    \item[(c1)] The standard benchmark set for CTL verification (small)~\cite{Cook2013a}.
    \item[(c2)] The standard benchmark set for CTL verification (industrial)~\cite{Cook2013a}.
    \item[(d)] The benchmark set of \muarith{} (i.e., \MuCLP{} restricted to integer arithmetic)~\cite{Kobayashi2019} which consists of some properties of integer arithmetic encoded in \muarith{} (Problems 1--6), linear-time temporal properties of first-order functional programs encoded by a translation in \cite{Kobayashi2019} (Problems 7--22), branching-time temporal properties (some are only expressible in CTL* or modal-$\mu$) of imperative programs encoded by a translation similar to one in \cite{Watanabe2019} (Problems 23--28).
    \item[(e)] The termination verification benchmark set for \function{}.\footnote{\url{https://www.di.ens.fr/\~urban/FuncTion.html}}
\end{itemize}
All experiments have been conducted on 3.1GHz Intel Xeon Platinum 8000 CPU and 32 GiB RAM with the time limit of 300 seconds.

\begin{figure*}[t]
\includegraphics[scale=0.45]{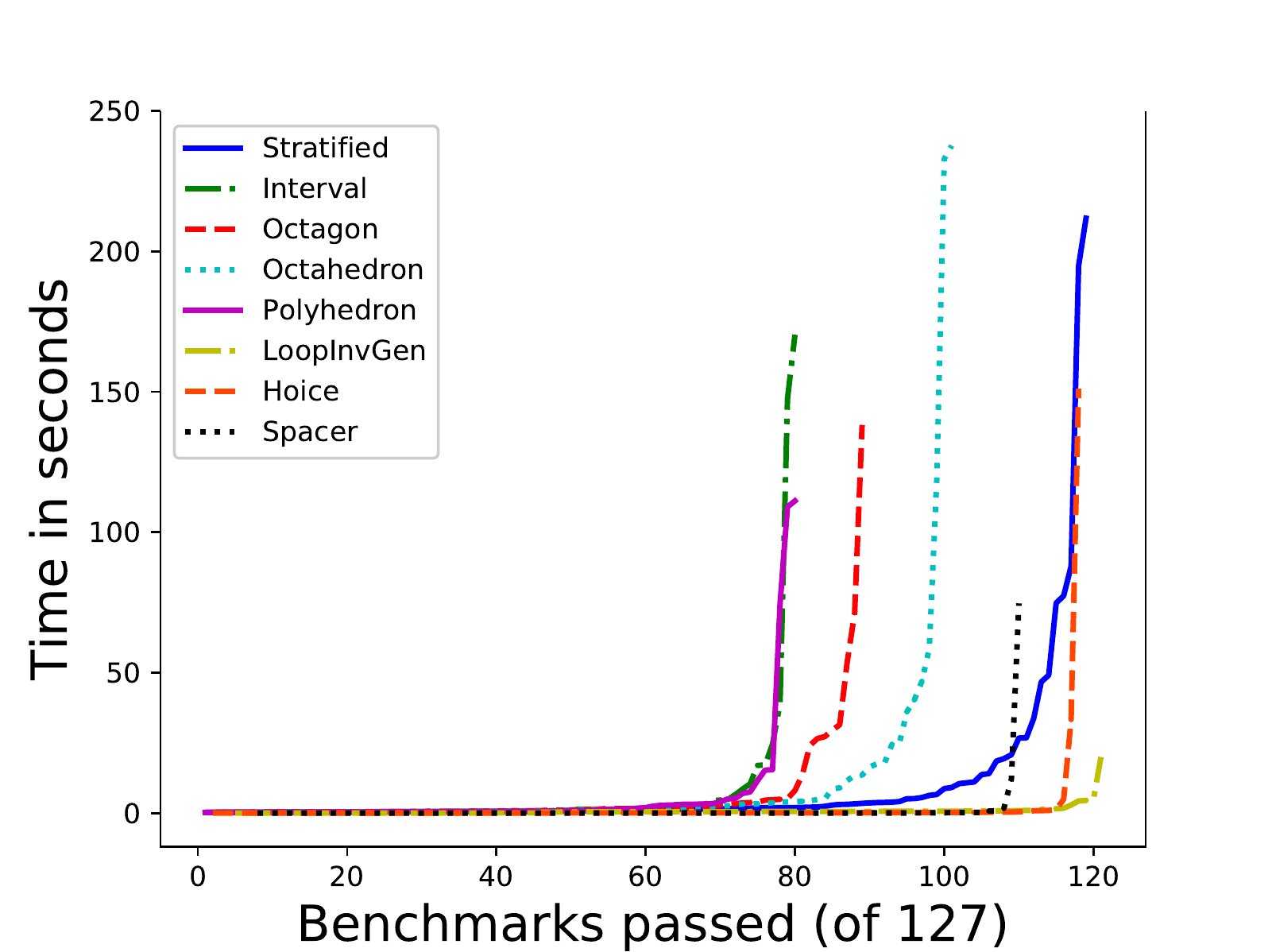}
\includegraphics[scale=0.24]{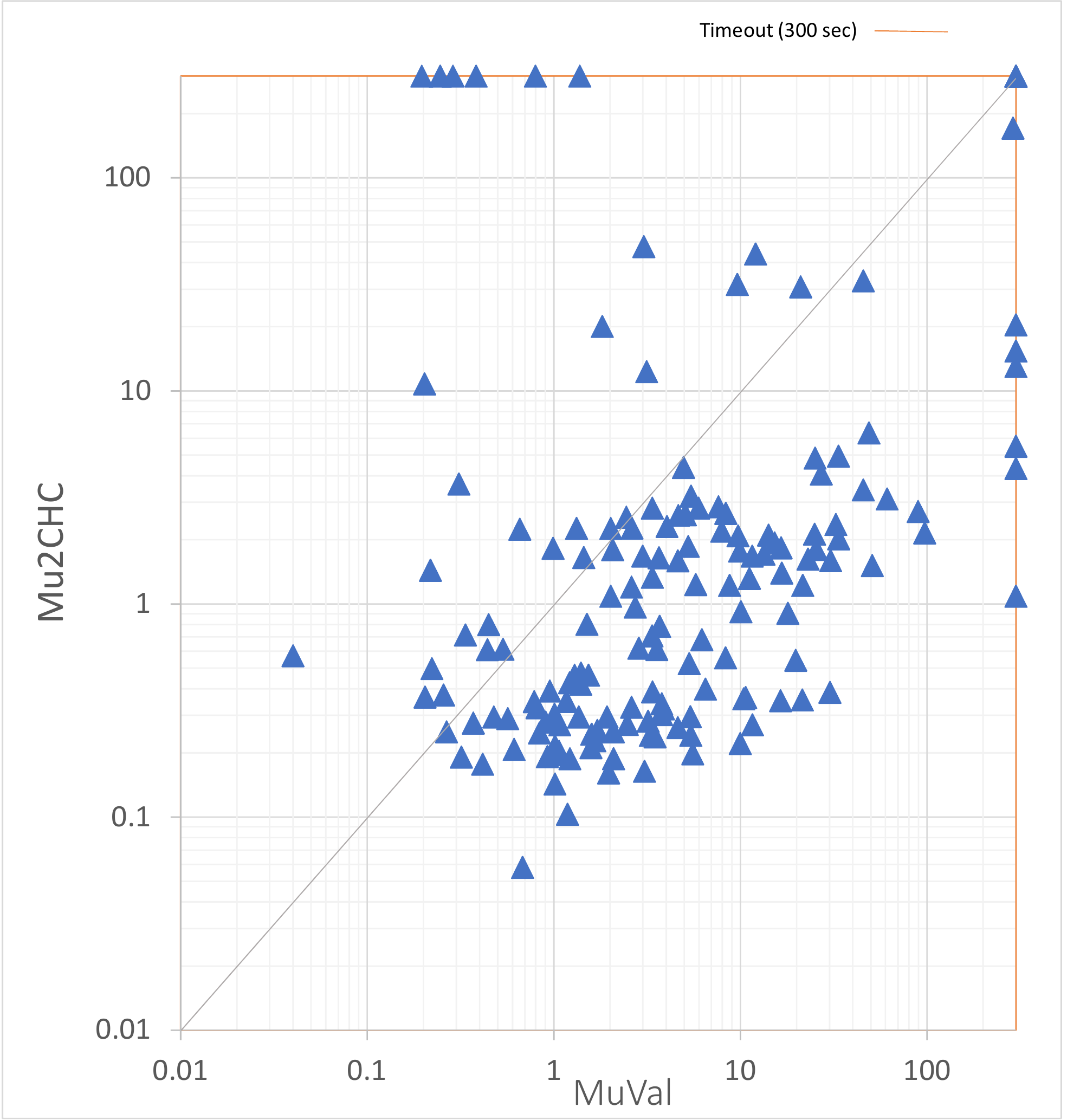}
\caption{Experimental results of \pcsat{} on the benchmarks from SyGuS-Comp 2018 (Inv Track) compared with \loopinvgen{}, \hoice{}, \spacer{} (left) and \muval{} on various \MuCLP{} benchmarks compared with \mutochc{}~\cite{Kobayashi2019} (right)}
\label{fig:exps}
\end{figure*}

The experimental results except (b) are summarized in Figure~\ref{fig:exps}.  The cactus plot (left) compares the results of \pcsat{} on (a) with those of \hoice{}, \spacer{}, and \loopinvgen{}.  For the number of solved instances, \pcsat{} obtained comparable results with \loopinvgen{}~\cite{Padhi2019}: \pcsat{} (denoted ``Stratified'') solved 113 SAT and 8 UNSAT instances while \loopinvgen{} solved 116 SAT and 5 UNSAT instances.  \pcsat{} obtained better results than the highly-tuned \CHCS{} solvers \hoice{} (109 SAT, 9 UNSAT, and 2 wrong answers) and \spacer{} (101 SAT and 9 UNSAT).  \pcsat{} however is often slower compared to the other mature tools.  This is partly because \pcsat{} does not use incremental SMT solving across CEGIS iterations and therefore becomes significantly slower as the number of example instances grows.  This inefficiency caused \pcsat{} to obtain suboptimal results on (b) the CHC-COMP benchmarks: \pcsat{} solved 97 SAT and 55 UNSAT instances while \hoice{} solved 123 SAT and 79 UNSAT, and \spacer{} solved 147 SAT and 117 UNSAT instances.  From our analysis of the failed runs, we found that
\pcsat{} often failed to solve \CHCS{} containing multiple Boolean variables.  This is because the current version of \pcsat{} naively generates $2^n$-copies of templates over integer variables for each Boolean valuation where $n$ is the number of Boolean variables in \CHCS{}.  We plan to design improved families of templates for Boolean variables.  Though it is rather out of the scope
of this paper, we believe this dramatically improves the experiment results because most benchmarks from (b) have multiple Boolean variables.
Also, we found that \pcsat{} is general but not well-tuned for proving the \emph{unsatisfiability} when applied to the subclass \CHCS{} of \PCSPWFFN{}.  We could exploit the restricted (i.e. Horn) form of constraints for efficiently finding a resolution derivation of the contradiction via SLD-resolution.

The cactus plot (left) also shows the trade-off between expressiveness and generalizability of templates.  Interval, Octagon, Octahedron, and Polyhedron are \pcsat{} restricted to use respective fixed predicate templates, and the plot shows that they obtained significantly worse results compared to \pcsat{} with stratified families of templates.  Also note that the results with the Polyhedron and the Interval templates are even worse than those of the Octahedron and the Octagon templates.  We believe that these results show that the Polyhedron templates suffer from the overfitting problem~\cite{Padhi2019} due to their high expressiveness, while the Interval templates suffer from their low expressiveness.

The scatter plot (right) in Figure~\ref{fig:exps} compares the results of \muval{} on (c1), (c2), (d), and (e) with those of \mutochc{}: \muval{} solved 76 VALID and 72 INVALID instances (out of 159 instances) and \mutochc{} solved 74 VALID and 74 INVALID instances.  \muval{} failed to solve 5 temporal verification benchmarks from (c2) and (d) that were solved by \mutochc{}.  We believe that this is because the highly-tuned invariant synthesis engine (i.e., \spacer{} and \hoice{}) used in \mutochc{} worked better for the benchmarks.
By contrast \mutochc{} failed to solve 5 termination verification benchmarks that were solved by \muval{}, which require synthesis of piecewise-defined and/or lexicographic affine ranking functions.  We believe that this shows a limitation of the \mutochc{} approach that separately synthesize termination arguments and inductive invariants, and cannot quickly feedback a failure of invariant synthesis to ranking function synthesis.

%% file: related.tex
The class of problems \PCSPWFFN{} that we have introduced in this paper is closely related to {\em existentially-quantified Horn clauses} (E-CHCs) introduced in \cite{Beyene2013}.  We conjecture that \PCSPWFFN{} and E-CHCs are inter-reducible, though it is not trivial to fill the gap, without changing the background theory, between our well-foundedness and their \emph{disjunctive} well-foundedness constraints and our function variables and their existentially-quantified heads.
We believe inter-reducibility is often a desirable feature: even though DFAs and regular expressions are inter-reducible, each format has its own benefits. In our case, for instance, having the direct support for general disjunctions in \PCSP{} can be advantageous compared to encoding them indirectly by existentials in E-CHCs.  In particular, general disjunctions can be handled by \pcsat{} without any additional twist, and can be used to completely encode branching-time temporal properties verification problems of imperative programs with finitely-bounded non-determinism,
for which existential quantifications in E-CHCs and function variables in \PCSPWFFN{} are probably overkill.
Also, the class of \PCSP{} without function variables and well-founded predicates is closed under negation like in \MuCLP{} (cf. Remark~\ref{rem:CHC_coCHC_MuCLP}).  Besides the logical beauty, the property is also useful in practice: we can mechanically compute the De Morgal dual of the given \CHCS{} and check the satisfiability of the primary and dual \CHCS{} in parallel or cooperatively.  Also, it is well known that well-founded relations used in \PCSPWFFN{}{} and disjunctively well-founded relations used in E-CHCs are both complete for termination (see \cite{Podelski2004a}) but have different benefits.  
To solve E-CHCs, \cite{Beyene2013} proposes a method called \ehsf{} which reduces the given E-CHCs to (ordinary) \CHCS{} by synthesizing candidate witnesses for existentially quantified variables iteratively in a counterexample-guided manner.  The generated \CHCS{} are then solved (possibly itself via a counterexample-guided iteration) by an off-the-shelf \CHCS{} solver. 
By contrast, our method, while also based on counterexample-guided iteration, reduces the problem to quantifier-free SMT solving by simultaneously synthesizing candidate invariants, well-founded relations, and quantifier witnesses.  We believe that there are two advantages to our approach.  One is that the simultaneous synthesis facilitates finding candidates that depend amongst each other, for instance, well-founded relations that depend on quantifier witnesses, by sharing useful information via faster feedbacks from synthesis failures.  Another advantage is that \PCSPWFFN{} can directly express {\em non}-Horn clauses whereas handling such clauses in E-CHCs would incur introducing additional existential quantifiers.

An extension of \CLP{} called co-Constraint Logic Programs (\COCLP{}) with mixed inductive and co-inductive predicates has been proposed in \cite{Saeedloei2012}.  Unlike our \MuCLP{}, \COCLP{} does not support mutually recursive inductive and co-inductive predicates which are necessary to directly express modal-$\mu$ temporal verification problems. %
Also related to our \MuCLP{} is \muarith{}~\cite{Lubarsky1993,Bradfield1999} which is a first-order fixpoint logic of integer arithmetic.  It has recently been applied to temporal property verification in~\cite{Kobayashi2019} where they present a method called \mutochc{} for checking the validity of formulas expressed in the logic.\footnote{Technically, their method works on hierarchical equation systems (HES) which is a reformulation of \muarith{}.} 
\mutochc{} works by reducing the problem to (ordinary) \CHCS{}.  This is done by conservatively approximating fixpoints by asserting some (symbolic) bound on their unfolding depths.  The resulting \CHCS{} are then solved by
an off-the-shelf \CHCS{} solver.  By contrast, our \muval{} reduces the problem
to \PCSPWFFN{} and therefore has the advantages of simultaneous synthesis remarked above.\footnote{In fact, \mutochc{} has no feedback from \CHCS{} solving to fixpoints approximation.}
In fact, this difference resulted in the better results of \muval{} on the termination verification benchmark set that requires synthesis of lexicographc and/or piecewise-defined ranking functions (recall discussion in Section~\ref{sec:eval}). 
And, the completeness of the reduction to \PCSPWFFN{} allows \muval{} to conclude the invalidity of the original \MuCLP{} from the unsatisfiability of the reduced \PCSPWFFN{} unlike \mutochc{}.
Also, \mutochc{} is specialized to integer arithmetic, for example, relying on that particular domain to encode existential quantifiers as fixpoints, as explained in Remark~\ref{rem:qelim}.  Generalizing their method to other theories (such as the theory of reals) may require non-trivial extensions.  By contrast,
\muval{} is designed for the full class of \MuCLP{} which can be seen as a generalization of \muarith{} to \emph{arbitrary first-order theories}.
However, we remark that both \mutochc{} and \ehsf{} have an advantage over our approach in that
they can utilize highly-tuned off-the-shelf \CHCS{} solvers.  Indeed, for this reason, we have noticed that our approach is often less efficient than theirs on (ordinary) \CHCS{} instances.

Our \PCSPWFFN{} solving technique generalizes a number of previous techniques developed for \CHCS{} solving and invariant/ranking function discovery.  Most closely related to our work are the data-driven approaches to solving \emph{subclasses of \CHCS{}} based on CEGIS~\cite{Solar-Lezama2006} combined with template-based synthesis via SMT solver~\cite{Sharma2013a,Garg2014}, greedy set covering with logic minimization~\cite{Sharma2013,Padhi2016}, decision tree learning~\cite{Krishna2015,Garg2016,Champion2018,Ezudheen2018,Zhu2018}, and grammar-based synthesis~\cite{Fedyukovich2018,Padhi2019}.  Our stratified CEGIS adopts the idea of stratified families of templates~\cite{Jhala2006,Terauchi2015}.  Our approach is similar in spirit to \cite{Padhi2019} but they use a stratified family of grammars instead and also do not use unsat cores for updating grammars.
The idea presented in \cite{Fedyukovich2018} of extracting grammars for enumerating ranking functions and recurrent sets and our idea of stratifying templates are orthogonal and could be better together.
Besides the data-driven approach, various \CHCS{} solving approaches have been proposed: counterexample-guided abstraction refinement and Craig interpolation~\cite{Unno2009,Hojjat2018}, generalized property directed reachability~\cite{Hoder2012,Komuravelli2014}, constraint specialization~\cite{Angelis2014a,Kafle2016}, and inductive theorem proving~\cite{Unno2017b}.
A number of existing techniques for program verification can be applied straightforwardly to invariant synthesis for linear \CHCS{} and ranking function synthesis.  Some use templates for invariants~\cite{Colon2003,Sankaranarayanan2004a} and ranking functions~\cite{Leike2014a} but many of them involve costly non-linear constraint solving.  RankFinder~\cite{Podelski2004} synthesizes linear ranking functions via linear constraint solving.  However, none of the above methods can be used to solve the full class of \PCSPWFFN{}.

%% file: conc.tex
We have introduced the class \MuCLP{} of constraint logic programs with arbitrarily nested inductive and co-inductive predicates and the class \PCSPWFFN{} of predicate constraint satisfaction problems that generalizes \CHCS{} with arbitrary clauses, function variables, and  well-foundedness constraints.  We have then established a program verification framework based on \MuCLP{} by showing that (1) \MuCLP{} can naturally encode various classes of verification problems, (2) the validity of \MuCLP{} can be reduced to the satisfiability of \PCSPWFFN{}, and (3) existing \CHCS{} solving and invariants/ranking function synthesis techniques can be adopted to \PCSPWFFN{} solving and further improved with the idea of stratified CEGIS for simultaneously achieving relative completeness (Theorem~\ref{thm:relcomp}) and practical effectiveness (Figure~\ref{fig:exps}, left).

Though we presented a sound and complete reduction from \MuCLP{} to \PCSPWFFN{} and the classes of \CHCS{} and \COCHCS{} correspond to fragments of \MuCLP{} as discussed in Remark~\ref{rem:CHC_coCHC_MuCLP}, any \MuCLP{} is reduced to the satisfiability of a \COCHCSWFFN{} that is a strict syntactic fragment of \PCSPWFFN{}
and recent semantic results based on the recursion theory~\cite{Tsukada2020} imply that the full class of \PCSPWFFN{} is strictly more expressive than \MuCLP{}, meaning that the full class of \PCSPWFFN{} is not necessary for the validity of \MuCLP{}.
It would thus be interesting to investigate the potential of the full class of \PCSPWFFN{} in practice and to find some (non-syntactic) restriction that would capture the full class of \MuCLP{}.

To further widen the applicability of our framework, we plan to extend our tools \muval{} and \pcsat{} to support other first-order theories beyond LIA/LRA such as arrays, algebraic data types (ADTs), and heaps~\cite{Duck2013}.  As far as the semantics of \MuCLP{} is concerned, there is no issue with the background theory being incomplete (i.e., undecidable).  However, the constraint solving method may require non-trivial extensions to support the above theories because it involves designing appropriate stratified families of templates.  For example, certificates (i.e., invariants, ranking functions, witnesses for quantifiers) over arrays often require quantifiers, and those over heaps and ADTs often require inductive predicates.
Future work also includes extensions of the framework to higher-order predicates and probabilities.  The former extension is useful for precisely analyzing higher-order recursive functions (cf. \HoCHC~\cite{Burn2018} and \HFL($\IntSet$)~\cite{Kobayashi2018,Watanabe2019}).  The latter extension is for reasoning about programs and systems that exhibit uncertain or probabilistic behaviors (cf. \PCHC~\cite{Albarghouthi2017b}).

%% file: main.bbl

%% file: apps.tex
We now consider a variety of verification problems, showing that each of them can be expressed as validity in \MuCLP{}.  Specifically, we discuss applications of \MuCLP{} to bisimulation and bisimilarity verification in Appendix~\ref{sec:apps_coind} and infinite state and infinite duration two player game solving in Appendix~\ref{sec:apps_game}.

\subsection{Bisimulation and Bisimilarity Verification}
\label{sec:apps_coind}
\input{apps_coind}

\subsection{Infinite State and Infinite Duration Games Solving}
\label{sec:apps_game}
\input{apps_game}

%% file: apps_coind.tex
Bisimulation and bisimilarity are a prototypical application of
greatest fixpoint and co-induction in computer
science~\cite{Sangiorgi2011}.  We show that the notions and various
problems thereof can be naturally expressed in our framework.

Given two LTS $\lts_1$ and $\lts_2$ with
$\labels = \labels_{\lts_1} = \labels_{\lts_2}$, the bisimilarity
relation $\bisim_{\lts_1,\lts_2}$ between $\lts_1$ and $\lts_2$ can be
defined in \MuCLP{} as follows:
\[
\begin{array}{l}
\bisim_{\lts_1,\lts_2}(\myseq{x}_1,\myseq{x}_2)  =_\nu\\
\qquad\begin{array}{ll}
\bigwedge_{\ell \in \labels} & \all{\myseq{y}_1} {\tuple{\myseq{x}_1} \lstep_{\lts_1} \tuple{\myseq{y}_1} \Rightarrow \exi{\myseq{y}_2}{\tuple{\myseq{x}_2} \lstep_{\lts_2} \tuple{\myseq{y}_2} \wedge \bisim_{\lts_1,\lts_2}(\myseq{y}_1,\myseq{y}_2)}}\\
    & \wedge \all{\myseq{y}_2}{\tuple{\myseq{x}_2} \lstep_{\lts_2} \tuple{\myseq{y}_2} \Rightarrow \exi{\myseq{y}_1}{\tuple{\myseq{x}_1} \lstep_{\lts_1} \tuple{\myseq{y}_1} \wedge \bisim_{\lts_1,\lts_2}(\myseq{y}_1,\myseq{y}_2)}}
    \end{array}
    \end{array}
\]
Note that the equation defines
$\bisim_{\lts_1,\lts_2}(\myseq{x}_1,\myseq{x}_2)$ as a greatest fixpoint.
A basic problem of interest in bisimulation is deciding whether two
(concrete) states, say $\myseq{n}_1 \in S_{\lts_1}$ and
$\myseq{n}_2 \in S_{\lts_2}$, are bisimilar.  This is expressed in our
logic by the formula $\bisim_{\lts_1,\lts_2}(\myseq{n}_1,\myseq{n}_2)$.
More generally, we may be interested in knowing if every pair of
states $\myseq{x_1} \in S_{\lts_1}$ and $\myseq{x_2} \in S_{\lts_2}$
satisfying $\phi(\myseq{x}_1,\myseq{x}_2)$ are bisimilar, where $\phi$ is
some property on pairs of states.  This can be expressed by the
formula
$\phi(\myseq{x}_1,\myseq{x}_2) \Rightarrow
\bisim_{\lts_1,\lts_2}(\myseq{x}_1,\myseq{x}_2)$.
  
Such queries are instances of checking if a formula is a {\em
  lower-bound} of a greatest fixpoint formula, and can be solved by
our constraint solving method described in Sections~\ref{sec:reduct}
and \ref{sec:csolve}.  As we shall show there, our technique for
solving such a constraint corresponds to the well-known technique of
proof by {\em co-induction}.

While co-induction can be used to prove lower-bounds of greatest
fixpoints, a different, new technique is required to prove their {\em
  upper-bounds}.  For instance, suppose that we wish to check if all
bisimilar pairs of states satisfy a certain property, say $\psi$.  The
query can be expressed in our logic by:
\( \bisim_{\lts_1,\lts_2}(\myseq{x}_1,\myseq{x}_2) \Rightarrow \psi(\myseq{x}_1,\myseq{x}_2)
\).
Solving such greatest-fixpoint upper-bound queries are beyond the
scope of previous methods.  Nonetheless, our method is able to solve
them by use of well-founded relations as we show in
Sections~\ref{sec:reduct} and \ref{sec:csolve}.

Next, we instantiate the above with a concrete instance.  Let us
consider a concrete LTS $\lts$ with labels $\labels_\lts = \{ +,-\}$, states
$S_\lts = \IntSet^2$, and the following transition relation:
\[
  \setlength{\arraycolsep}{4pt}
  \begin{array}{ll}
    \tuple{x,y} \step{+} \tuple{x+1,y} & \text{if }x+1 \leq y \\
    \tuple{x,y} \step{-} \tuple{x-1,y} & \text{if }x-1 \geq y
  \end{array}
\]
Let us consider $\bisim_{\lts,\lts}$, that is, we consider the
bisimulation relating two states of the same system $\lts$.  We may
then check if two states, for instance $(0,1) \in S_\lts$ and
$(1,2) \in S_\lts$, are bisimilar by proving if
$\bisim_{\lts,\lts}(0,1,1,2)$ is true.  In this case, our method is
able to do the proof by synthesizing the co-inductive invariant
$\uapprox{\bisim_{\lts,\lts}}(x_1,y_1,x_2,y_2) \defeq x_1=0 \land x_2=1 \land x_3=1 \land x_4=2 \lor
x_1=1 \land x_2=1 \land x_3=2 \land x_4=2$.
Our method can also prove a more general property that any states of $\lts$
such that $y-x$ is the same are bisimilar, by synthesizing the co-inductive invariant $\uapprox{\bisim_{\lts,\lts}}(x_1,y_1,x_2,y_2) \defeq y_1-x_1=y_2-x_2$.

Next, suppose that we wish to prove that every pair of bisimilar
states $(x_1,y_1) \in S_\lts$ and $(x_2,y_2) \in S_\lts$ satisfies
$y_1-x_1 = y_2-x_2$.
That is, every bisimilar states of $\lts$ have equal direction and distance from $x$
to $y$.  The query can be expressed in our logic by the following
formula:
\[
  \bisim_{\lts,\lts}(x_1,y_1,x_2,y_2) \Rightarrow y_1-x_1 = y_2-x_2
\]
which is equivalent to
\[
  y_1-x_1 \neq y_2-x_2 \Rightarrow \bisim_{\lts,\lts}^\neg(x_1,y_1,x_2,y_2)
\]
where $\bisim_{\lts,\lts}^\neg$ is the de Morgan dual of $\bisim_{\lts,\lts}$ defined by:
\[
\begin{array}{l}
    \bisim_{\lts_1,\lts_2}^\neg(\myseq{x}_1,\myseq{x}_2)  =_\mu \\
   \qquad \begin{array}{ll} 
                              \bigvee_{\ell \in \labels} & \exi{\myseq{y}_1}{\tuple{\myseq{x}_1} \lstep_{\lts_1} \tuple{\myseq{y}_1} \land (\all{\myseq{y}_2}{\tuple{\myseq{x}_2} \lstep_{\lts_2} \tuple{\myseq{y}_2} \imply \bisim_{\lts_1,\lts_2}^\neg(\myseq{y}_1,\myseq{y}_2)})}\\
    &  \lor \exi{\myseq{y}_2}{\tuple{\myseq{x}_2} \lstep_{\lts_2} \tuple{\myseq{y}_2} \land (\all{\myseq{y}_1}{\tuple{\myseq{x}_1} \lstep_{\lts_1} \tuple{\myseq{y}_1} \imply \bisim_{\lts_1,\lts_2}(\myseq{y}_1,\myseq{y}_2)})}
  \end{array}
  \end{array}
\]
As remarked above, such a ``property checking'' query on greatest
fixpoints can be, as a result, handled by our method by using well-founded
relations.  Here, our method synthesizes the inductive invariant
$\uapprox{\bisim_{\lts,\lts}^\neg}(x_1,y_1,x_2,y_2) \defeq y_1-x_1 \neq y_2-x_2$ and the well-founded relation
$\wfp{\bisim_{\lts,\lts}^\neg}(x_1,y_1,x_2,y_2,x_1',y_1',x_2',y_2') \defeq r(x_1,y_1,x_2,y_2) >
r(x_1',y_1',x_2',y_2')$ where
$r(x_1,y_1,x_2,y_2) = |(y_1-x_1) + (y_2-x_2)|$.

%% file: apps_game.tex
{\em Two-player turn-based infinite-duration games} are games in which
two players take turns in moving a token along the edges of a graph.
A player wins if the (infinite) sequence of nodes visited by the token
satisfies a certain condition.  Classically, the games are played over
a finite graph representing the state transition diagram of a
finite-state transition system, and there is a rich body of work
relating such games to the verification and synthesis of finite state
systems~\cite{Graedel2002}.
For instance, in the synthesis of reactive
systems~\cite{Buchi1969,Pnueli1989,Thomas1995},
a game with two players, {\sf Sys} and {\sf Env}, is considered over a
graph with edges from one player's node to the other player's node.
The edges from {\sf Sys}'s nodes describe the possible (one-step)
execution choices of the system to be synthesized and those from {\sf
Env}'s nodes describe the possible external inputs to the system.  The
goal of {\sf Sys} is to satisfy the given specification (given, e.g.,
by a temporal logic formula) whereas the goal of {\sf Env} is to
violate it.  The desired system is realizable if and only if {\sf Sys}
has a winning strategy.

Recently, the line of work has been extended to {\em infinite-state}
systems with which one can express the verification and synthesis
problems for infinite-state systems~\cite{Beyene2014,Farzan2018}.
We show that our framework is expressive enough to express such
infinite-state infinite-duration games.  Following the
literature~\cite{Beyene2014,Farzan2018}, we consider three classes of
games: {\em Safety games}, {\em Reachability games}, and {\em LTL
  games}.\footnote{``LTL games'' is a misnomer as the games actually go beyond LTL properties.  \cite{Farzan2018} also considers another
  class of games called {\em Satisfiability games} which is only
  finite duration and therefore is omitted from our discussion.}

Each game is played over a graph formed by a LTS.  Specifically, we
consider a LTS of the form $\lts = (S,T,\labels_\aplayer \cup
\labels_\eplayer)$ where $\labels_\aplayer \cap \labels_\eplayer = \emptyset$, that is, the
labels are partitioned into $\labels_\aplayer$ and $\labels_\eplayer$.
For each $\ell \in \labels_\eplayer$ (resp.~$\ell \in \labels_\aplayer$), a
transition $s \lstep s'$ denotes $\eplayer$ player's (resp.~$\aplayer$
player's) move from node $s$ to node $s'$.  We write $s \step{\eplayer} s'$
(resp.~$s \step{\aplayer} s'$) when there exists $\ell \in \labels_\eplayer$
(resp.~$\ell \in \labels_\aplayer$) such that $s \lstep s'$.
Below, we describe each class of games and our encoding of them in
\MuCLP{}.  For simplicity, we assume that each game starts with
$\aplayer$'s turn.

\subsubsection{Safety games}
\label{sec:apps_game_safe}
In a {\em safety game}, we are given predicates $\phi_{\it
init}(\myseq{x})$ and $\phi_{\it safe}(\myseq{x})$.  The $\eplayer$ player
wins the game if only states satisfying $\phi_{\it safe}(\myseq{x})$ are
visited along any sequence of plays starting from any state satisfying
$\phi_{\it init}(\myseq{x})$.
The game can be expressed in \MuCLP{} by defining the greatest
fixpoint predicate $\sg(\myseq{x})$ as follows:
\[
\sg(\myseq{x}) =_\nu \phi_{\it safe}(\myseq{x}) \wedge \forall\myseq{y}. \tuple{\myseq{x}} \step{\aplayer} \tuple{\myseq{y}} \Rightarrow \phi_{\it safe}(\myseq{y}) \wedge \exists\myseq{z}. \tuple{\myseq{y}} \step{\eplayer} \tuple{\myseq{z}} \wedge \sg(\myseq{z}).
\]
Then, $\eplayer$ has a winning strategy if and only if $\phi_{\it
init}(\myseq{x}) \Rightarrow \sg(\myseq{x})$ is valid.  The correctness of
the encoding can be readily seen by observing that $\sg(\myseq{x})$
describes exactly the set of states from which $\eplayer$ can force
the plays to stay in the states satisfying $\phi_{\it safe}(\myseq{x})$.

\subsubsection{Reachability games}
\label{sec:apps_game_reach}
In a {\em reachability game}, we are given predicates
$\phi_{\it init}(\myseq{x})$ and $\phi_{\it reach}(\myseq{x})$.  The
$\eplayer$ player wins the game if for any play starting from a state
satisfying $\phi_{\it init}(\myseq{x})$, a state satisfying
$\phi_{\it reach}(\myseq{x})$ is eventually visited.  As clear from the
definition, reachability games are the dual of safety games.  That is,
a safety game with the objective $\phi_{\it safe}(\myseq{x})$ is won by
$\eplayer$ if and only if $\aplayer$ wins the reachability game with
the objective
$\phi_{\it reach}(\myseq{x}) = \neg \phi_{\it safe}(\myseq{x})$ on the
same graph but with the players' edge sets swapped.
The game can be expressed in \MuCLP{} by defining the least
fixpoint predicate $\rg(\myseq{x})$ as follows:
\[
\rg(\myseq{x}) =_{\mu} \phi_{\it reach}(\myseq{x}) \vee \forall\myseq{y}. \tuple{\myseq{x}} \step{\aplayer} \tuple{\myseq{y}} \Rightarrow \phi_{\it reach}(\myseq{y}) \vee \exists\myseq{z}. \tuple{\myseq{y}} \step{\eplayer} \tuple{\myseq{z}} \wedge \rg(\myseq{z}).
\]
Then, $\eplayer$ has a winning strategy if and only if
$\phi_{\it init}(\myseq{x}) \Rightarrow \rg(\myseq{x})$ is valid.  The
correctness of the encoding can be readily seen by observing that
$\rg(\myseq{x})$ describes exactly the set of states from which
$\eplayer$ can force the plays to eventually reach a state satisfying
$\phi_{\it reach}(\myseq{x})$.

Our \MuCLP{} formulations of safety games and reachability games show
a striking resemblance, reflecting the inherent duality of the two
classes of games.  This is in contrast to their formulations in
{\em existentially-quantified Horn
  clauses}~\cite{Beyene2013,Beyene2014} that used rather different
encodings for the two classes.

\subsubsection{LTL games}
In a {\em LTL game}, we are given a predicate
$\phi_{\it init}(\myseq{x})$ and a B\"{u}chi automaton $\buchi$ such
that the label set of $\buchi$ is
$\labels_\eplayer \cup \labels_\aplayer$
(cf.~Section~\ref{sec:apps_temp} for the definition of B\"{u}chi
automaton).  The $\eplayer$ player wins the game if for any play
starting from a state satisfying $\phi_{\it init}(\myseq{x})$, the
infinite sequence of labels of the play is accepted by $\buchi$.
The game can be expressed in \MuCLP{} by defining the mutually-recursive
least-and-greatest fixpoint predicates $\ltlg_{q,\alpha}(\myseq{x})$ 
for each $q \in Q$ and $\alpha \in \{\mu,\nu\}$:
\[
\ltlg_{q,\alpha}(\myseq{x}) =_\alpha \bigwedge_{\ell \in \labels_\aplayer} \forall\myseq{y}.\tuple{\myseq{x}} \lstep \tuple{\myseq{y}} \Rightarrow \bigvee_{q' \in \delta(q,\ell)}\bigvee_{\begin{subarray}{c} \ell' \in \labels_\eplayer \\ q'' \in \delta(q',\ell') \end{subarray}} \exists\myseq{z}.\tuple{\myseq{y}} \step{\ell'} \tuple{\myseq{z}} \wedge \ltlg_{q'',\alpha(q',q'')}(\myseq{z}).
\]
Here, $\alpha(q_1,q_2) = \nu$ if $q_1 \in F$ or $q_2 \in F$, and
$\alpha(q_1,q_2) = \mu$ otherwise.  Then, $\eplayer$ has a winning
strategy if and only if
$\phi_{\it init}(\myseq{x}) \Rightarrow \ltlg_{q_{\it
    init},\nu}(\myseq{x})$ is valid.  The correctness of the
construction follows from an argument similar to that of the linear
temporal property verification reduction shown in
Section~\ref{sec:apps_temp}.

Here is an example of a simple LTL game for the property
$\textsf{GF}(\textsf{restore})$.  The game consists of a single
integer variable $x$ whose value is initially 0.  The LTS is defined
with the label sets
$\labels_\eplayer = \{ \textsf{restore}, \textsf{incr}, \textsf{decr}
\}$ and $\labels_\aplayer = \{ \textsf{break}, \textsf{skip} \}$, and
the transition relation is shown in the left column below:

\begin{tabular}{ll}
  \begin{minipage}{.6\linewidth}
    \begin{itemize}
    \item $\tuple{x} \step{\textsf{restore}} \tuple{x'}$ if $x=x'=0$
    \item $\tuple{x} \step{\textsf{incr}} \tuple{x'}$ if $x' = x + 1$.
    \item $\tuple{x} \step{\textsf{decr}} \tuple{x'}$ if $x' = x - 1$.
    \item $\tuple{x} \step{\textsf{break}} \tuple{x'}$ if $x=0$ and $x' \neq 0$.
    \item $\tuple{x} \step{\textsf{skip}} \tuple{x'}$ if $x=x'$.
    \end{itemize}
  \end{minipage}
  &
    \begin{minipage}{.4\linewidth}
      \xymatrix@!C=2pc{%
        *+<1pc>[o][F=]{q_0} \ar@(l,l)^<{} \ar@(lu,ur)[]^{\textsf{restore}} \ar@/_/[r]_{*} 
        & *+<1pc>[o][F-]{q_1}  \ar@/_/[l]_{\textsf{restore}} \ar@(lu,ur)[]^{*}
        }
    \end{minipage}
\end{tabular}
\newline
A B\"{u}chi automaton expressing the property $\textsf{GF}(\textsf{restore})$
is shown in the right column.

The winning strategy is obvious: whenever player $\aplayer$ \textsf{break}s away from $q_0$ by randomly assigning to $x$, $\eplayer$ must either \textsf{incr} or \textsf{decr} to get $x$ back to 0 and then \textsf{restore}.
Following the encoding of $\ltlg_{q,\alpha}$ above, it is straight forward to define this game in \MuCLP{}. The encoding will be as follows:
\[
  \begin{array}{l}
    \ltlg_{q,\alpha}(x) =_\alpha (\forall x'.\ (x=0 \wedge x' \neq 0) \Rightarrow \phi(x')) \wedge (\forall x'.\ x' = x \Rightarrow \phi(x')) \\
    \phi(x') = 
    \bigvee \left\{
    \begin{array}{l}
      \exists x''. x'' = x'-1 \wedge \ltlg_{q_1,\mu}(x'') \\
      \exists x''. x'' = x'+1 \wedge \ltlg_{q_1,\mu}(x'') \\
      \exists x''. x'' = x' = 0 \wedge \ltlg_{q_0,\nu}(x'')
    \end{array} \right.
  \end{array}
\]
where $q \in \{q_0,q_1 \}$ and $\alpha \in \{ \mu, \nu \}$.
The universally quantified actions pertain to the $\aplayer$ player performing a \textsf{break} or \textsf{skip} action. The existentially quantified actions pertain to the $\eplayer$ player performing a \textsf{decr}, \textsf{incr}, or \textsf{restore} action.

 \subsubsection{Cinderella-Stepmother game}

 \newcommand{\cd}{\textsf{Cinderella}}
 \newcommand{\sm}{\textsf{Stepmother}}
 \newcommand{\nnrat}{\mathbb{Q}_{\geq 0}}

 \newcommand{\overflowlab}{\mathsf{ov}}
 \newcommand{\smlab}{\mathsf{sm}}
 \newcommand{\cdlab}{\mathsf{cd}}
 \newcommand{\smlabels}{\labels_{\mathsf{SM}}}
 \newcommand{\cdlabels}{\labels_{\mathsf{CD}}}

 As a concrete example of the three classes of games, let us consider
 the Cinderella-Stepmother
 game~\cite{Hurkens2011,Bodlaender2012},
 which is also used as examples in \cite{Beyene2014,Farzan2018}.  The
 game comprises five buckets of water arranged in a circle.  Each
 bucket can hold some constant $c$ amount of water.  The two players,
 \cd{} and \sm{}, take turns emptying and filling the buckets.  In each
 of her turns, \sm{} brings 1 unit of additional water and distributes it
 among the five buckets.  In turn, \cd{} chooses two adjacent buckets
 and empties them.  \cd{} wins if none of the buckets ever overflow.
 It is known that \cd{} has a winning strategy exactly when $c > 2$.
 For instance, when $c < 1$, it is easy to see that \sm{} wins in one
 round by pouring the entire additional water to a single bucket.
 Also, when $c \geq 3$, it is easy to see that \cd{} can win by
 adopting the {\em round-robin} strategy whereby she goes around the
 circle and in each round empties two buckets that are adjacent to the
 buckets that were emptied in the previous round.  However, as remarked
 in \cite{Beyene2014,Farzan2018}, synthesizing a winning strategy (for
 \cd{} or \sm{}) when $1 \leq c < 3$ is non-trivial.

 We formalize the game in our framework.  The state space of the game
 is $S = \nnrat^5$ where $\nnrat$ is the set of non-negative rational
 numbers.  Each $(b_0,b_1,b_2,b_3,b_4) \in S$ represents the state of
 the five buckets with $b_i$ being the amount of water in the $i$-th
 bucket.  The set of labels for \sm{} is
 $\smlabels = \{ \overflowlab,\smlab \}$ where $\overflowlab$ indicates
 that a bucket is overflowing after \sm{} has made the move.  The set
 of labels for \cd{} is the singleton set $\cdlabels= \{\cdlab \}$.
 Let us write $[0,4]$ for the set $\{0,1,2,3,4 \}$.  The transition
 relation is defined by:
 \begin{itemize}
 \item $(b_0,b_1,b_2,b_3,b_4) \step{\smlab} (b_0',b_1',b_2',b_3',b_4')$ if
   $1 + \sum_{i \in [0,4]} b_i = \sum_{i \in [0,4]} b_i'$ and $b_i \leq b_i' \leq c$ for each $i \in [0,4]$.
 \item $(b_0,b_1,b_2,b_3,b_4) \step{\overflowlab} (b_0',b_1',b_2',b_3',b_4')$ if
   $1 + \sum_{i \in [0,4]} b_i = \sum_{i \in [0,4]} b_i'$, $b_i \leq b_i'$ for each $i \in [0,4]$, and there exists $i \in [0,4]$ such that $b_i' > c$.
 \item $(b_0,b_1,b_2,b_3,b_4) \step{\cdlab} (b_0',b_1',b_2',b_3',b_4')$ if there exists $i \in [0,4]$ such that $b_i' = b_{(i+1)\% 5}' = 0$ and $b_j = b_j'$ for each
   $j \in [0,4] \setminus \{i,(i+1)\% 5\}$.
 \end{itemize}
 The set of initial states is described by $\phi_{\it init}(b_0,b_1,b_2,b_3,b_4) \defeq \bigwedge_{i \in [0,4]} b_i = 0$, that is, the buckets are initially all empty.

 For \cd{}, the game can be formalized as a safety game where
 $\labels_\eplayer = \cdlabels$, $\labels_\aplayer = \smlabels$, and the
 safety objective is
 $\phi_{\it safe}(b_0,b_1,b_2,b_3,b_4) \defeq \bigwedge_{i \in [0,4]} b_i
 \leq c$.
 Dually, for \sm{}, the game can be formalized as a reachability game
 where $\labels_\eplayer = \smlabels$, $\labels_\aplayer = \cdlabels$,
 and the reachability objective is
 $\phi_{\it reach}(b_0,b_1,b_2,b_3,b_4) \defeq \neg \phi_{\it
   safe}(b_0,b_1,b_2,b_3,b_4) = \bigvee_{i \in [0,4]} b_i > c$.

 \begin{figure*}[t]
   \[
     \setlength\arraycolsep{2em}
     \begin{array}{cc}
       \xymatrix@!C=2pc{
       *+<1pc>[o][F-]{} \ar@(l,l)^<{} \ar@(lu,ur)[]^{*} \ar[r]^{*} 
       & *+<1pc>[o][F=]{}  \ar@(lu,ur)[]^{\smlab,\cdlab}
         }
       &
         \xymatrix@!C=2pc{%
         *+<1pc>[o][F=]{} \ar@(l,l)^<{} \ar@(lu,ur)[]^{\smlab,\cdlab} \ar@/_/[r]_{*} 
       & *+<1pc>[o][F-]{}  \ar@/_/[l]_{\smlab,\cdlab} \ar@(lu,ur)[]^{*}
         }
        
       \\
       \buchi_1 & \buchi_2
     \end{array}
   \]
   \caption{B\"{u}chi automata for Cinderella-Stepmother LTL games.}
   \label{fig:cdsbuchi}
 \end{figure*}
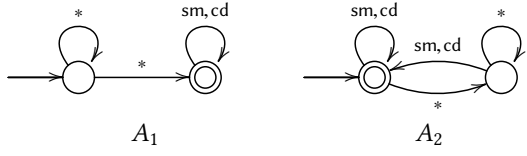

 As in \cite{Beyene2014}, let us also consider variants of the game
 with LTL objectives.  For instance, consider the B\"{u}chi automata
 $\buchi_1$ and $\buchi_2$ shown in Fig.~\ref{fig:cdsbuchi}.
 $\buchi_1$ corresponds to the LTL formula
 $\mathsf{FG}(\neg \overflowlab)$.  That is, it accepts exactly
 the plays in which overflows happen only finitely often.  By contrast,
 $\buchi_2$ corresponds to the LTL formula
 $\mathsf{GF}(\neg \overflowlab)$ and it accepts exactly the
 plays in which buckets are in a non-overflowing state infinitely often.
 As also remarked in \cite{Beyene2014}, the automata are examples of
 {\em weakened} objectives for \cd{} which may allow her to win the game more
 often.  Using $\buchi_1$ or $\buchi_2$ as the objective B\"{u}chi
 automaton and letting $\labels_\eplayer = \cdlabels$ and
 $\labels_\aplayer = \smlabels$, our framework is able to model the
 weakened variants as LTL games.